\documentclass[iop,apj,12pt]{emulateapj}
\usepackage{times}
\usepackage[normalem]{ulem}
\usepackage{epsfig}
\usepackage{natbib}
\usepackage{amsfonts}
\usepackage{amsmath}
\usepackage{multirow}
\usepackage{enumerate}
\usepackage[usenames]{color}
\usepackage[plainpages=false, colorlinks=true, anchorcolor=blue, linkcolor=blue, citecolor=blue, bookmarks=false]{hyperref}
\newcommand{\fheat}{$f_\textrm{heat}$}

\def\sun{_\odot}

\def\msun{M$_\odot$}

\def\Dwa{$\,$\uppercase\expandafter{\romannumeral5}$\,$}

\def\sless{\lower2pt\hbox{$\buildrel {\scriptstyle <}
   \over {\scriptstyle\sim}$}}

\def\sgreat{\lower2pt\hbox{$\buildrel {\scriptstyle >}
   \over {\scriptstyle\sim}$}}
\def\sharpnull#1{}

\newcommand{\subdate}{2014 August 6}
\newcommand{\shortauth}{Couch \& Ott}
\newcommand{\slugcom}{Submitted to ApJ on \subdate}
\slugcomment{\slugcom}

\lefthead{\sc \footnotesize \slugcom \hfill \shortauth}
\righthead{\sc \footnotesize \slugcom \hfill \shortauth}

\begin{document}

\title{The Role of Turbulence in Neutrino-Driven Core-Collapse Supernova Explosions}


\author{Sean M. Couch\altaffilmark{1,2,4}}
\author{Christian D. Ott\altaffilmark{2,3,5}}
  \altaffiltext{1}{Flash Center for Computational Science, Department of Astronomy \& Astrophysics, University of Chicago, Chicago, IL, 60637, smc@flash.uchichago.edu}
  \altaffiltext{2}{TAPIR, Walter Burke Institute for Theoretical Physics, MC 350-17,
  California Institute of Technology, Pasadena, CA 91125, USA, 
  cott@tapir.caltech.edu}
\altaffiltext{3}{Kavli Institute for the Physics and
 Mathematics of the Universe (Kavli IPMU WPI), The University of Tokyo, Kashiwa, Japan}
\altaffiltext{4}{Hubble Fellow}
\altaffiltext{5}{Alfred P. Sloan Research Fellow}

\begin{abstract}

The neutrino-heated ``gain layer'' immediately behind the stalled shock in a core-collapse supernova is unstable to high-Reynolds-number turbulent convection. 
We carry out and analyze a new set of 19 high-resolution three-dimensional (3D) simulations with a three-species neutrino leakage/heating scheme and compare with spherically-symmetric (1D) and axisymmetric (2D) simulations carried out with the same methods. 
We study the postbounce supernova evolution in a $15$-$M_\odot$ progenitor star and vary the local neutrino heating rate, the magnitude and spatial dependence of asphericity from convective burning in the Si/O shell, and spatial resolution. 
Our simulations suggest that there is a direct correlation between the strength of turbulence in the gain layer and the susceptability to explosion.
2D and 3D simulations explode at much lower neutrino heating rates than 1D simulations. 
This is commonly explained by the fact that nonradial dynamics allows accreting material to stay longer in the gain layer. 
We show that this explanation is incomplete.
Our results indicate that the effective turbulent ram pressure exerted on the shock plays a crucial role by allowing multi-D models to explode at a lower postshock thermal pressure and thus with less neutrino heating than 1D models.
We connect the turbulent ram pressure with turbulent energy at large scales and in this way explain why 2D simulations are erroneously exploding more easily than 3D simulations.


\end{abstract}


\section{Introduction}
\label{sec:intro}
The core-collapse supernova mechanism is a resilient problem. 
On the one hand, we know with certainty that core-collapse supernovae arise from the deaths of stars with zero-age main-sequence masses above about 8 \msun\ \citep[e.g.,][]{smartt:09b}, yet decades of theoretical investigation have failed to uncover a robust mechanism that turns these stars inside-out in energetic, luminous displays \citep{arnett:66, colgate:66, bethewilson:85, janka:12a, burrows:13a}.  At the end of their lives, such massive stars form a degenerate iron core that collapses once it reaches its effective Chandrasekhar mass.
Collapse of the inner core is halted only once supra-nuclear densities are attained, at which point the repulsive core of the strong nuclear force stiffens the nuclear equation of state (EOS). This results in core bounce and a strong shock is launched into the still collapsing outer core.
This ``bounce'' shock, however, is not strong enough to blow up the star.  
It stalls at a typical radius of $\sim$$150\,\mathrm{km}$ within only tens of milliseconds of bounce and turns into an accretion shock.
This is due to energy losses to dissociation of iron-group nuclei at the shock 
and to neutrinos that stream away from the now semi-transparent postshock region.
The driving question for decades in core-collapse supernova theory has been what revives the stalled shock, allowing it to drive robust explosions that resemble observed core-collapse supernovae?

Core collapse and the subsequent cooling of the protoneutron star liberate gravitational binding energy of some $\sim$$3\times10^{53}\,\mathrm{erg}$.
This is $\gtrsim$$100$\,times the typical explosion energy of observed core-collapse supernovae, and the vast majority of this energy is released in neutrinos during the first tens of seconds following collapse \citep[e.g.,][]{burrows:86}. 
This prediction of core-collapse supernova theory was impressively confirmed by the observation of neutrinos from SN 1987A \citep{hirata:87,bionta:87}. 
Any core-collapse supernova explosion mechanism must tap the binding energy reservoir and convert the necessary fraction to power the explosion. Since neutrinos are the primary energy transporters, investigation in recent decades has focussed on the  (delayed) neutrino mechanism \citep{bethewilson:85}. 
In the simplest picture, the neutrino mechanism relies on neutrino absorption increasing the thermal pressure in the immediate postshock region (the ``gain region,'' where absorption dominates over emission). 
This offsets the pressure balance with the ram pressure of the accreting outer core, leading to a runaway explosion. 
Detailed spherically-symmetric (1D) simulations have shown that the neutrino mechanism generally fails to produce explosions \citep[e.g.,][]{ramppjanka:00,liebendoerfer:03,thompson:03,sumiyoshi:05} for all but the lowest-mass progenitors \citep{kitaura:06}.  Once faster computers and more sophisticated simulation codes made high-fidelity axisymmetric (2D) simulations feasible, it was quickly realized that non-spherically-symmetric phenomena such as protoneutron star convection (e.g.,\citealt{burrows:93b,keil:96,mezzacappa:98b,dessart:06pns}), neutrino-driven convection in the postshock region between protoneutron star and shock \citep{herant:94,bhf:95,janka:96}, and the standing accretion shock instability \citep[SASI,][]{blondin:03,scheck:08} aid the neutrino mechanism. 
Successful, albeit under-energetic, explosions have been obtained in certain 2D cases \citep{marek:09, mueller:12a, mueller:12b, bruenn:13}, but not in others \citep[e.g.,][]{dolence:14,ott:08}. 

The success of 2D over 1D raised hopes that in 3D robust explosions would be found. 
The handful of 3D simulations including sophisticated neutrino transport have somewhat dashed those hopes, however, as they fail to explode, even for progenitors that explode in 2D \citep{hanke:13, tamborra:14}.  
As argued by \citet{couch:13b} and \citet{couch:14a}, this result is, regrettably, to be expected since 2D simulations are artificially prone to explosions.
This is the case because 2D cannot adequately model the essentially non-axisymmetric dynamics of core-collapse supernova, in particular SASI, convection, and turbulence.
The forced symmetry of 2D simulations exaggerates the positive influence of these dynamics on shock expansion (\citealt{hanke:12, couch:13b, takiwaki:14a, couch:14a}, but see \citealt{dolence:13,handy:14} for an alternative view on the 2D vs.\ 3D question).

Thus, we are left with the same, old question: what ingredient is missing from the most sophisticated 3D core-collapse supernova simulations that would trigger robust explosions?
In \citet{couch:13d}, we pointed out that significant progenitor asphericity, as is expected from multi-dimensional simulations of precollapse nuclear shell burning in core-collapse supernova progenitors \citep[e.g.,][]{meakin:07b, arnett:11a, couch:14c}, can trigger explosion in a 3D model that otherwise fails to explode with spherically-symmetric initial conditions.  
We argued that the precollapse velocity fluctuations we imposed onto the 1D progenitor models, once accreted through the shock, substantially enhanced the strength of nonradial motions in the gain region.
We surmised that this enhanced nonradial dynamics increased the typical time a fluid parcel entering through the shock spent in the gain region (the ``dwell time'') before it was advected to smaller radii where cooling dominates. In turn, the neutrino heating efficiency increased, facilitating explosion.

Similar arguments are made and universally accepted for the generally favorable effect of multi-dimensional dynamics on the neutrino mechanism \citep[e.g.,][]{bhf:95,murphy:08,marek:09,dolence:13,mueller:12a,bruenn:13,takiwaki:14a}:  multi-dimensional motions increase the dwell time of material in the gain region, enhance the efficiency of neutrino heating, and, in simulations using temporally and spatially fixed input neutrino luminosities (the so-called ``light bulb'' approximation), lead to explosion at lower critical luminosities than in 1D. 
Implicit in this dwell-time argument is that the matter behind the shock attains roughly the same thermal state as in the critical 1D case.
In other words, the explosion criteria based on the thermal state of the postshock matter \citep[e.g.,][]{burrows:93, pejcha:12a} are unchanged, the postshock matter is simply more efficient at achieving the critical thermal state for a given neutrino luminosity.
This implicit assumption for the relative ease of multi-dimensional explosions has gone largely unproven in the core-collapse supernova literature.

\emph{While it is true that nonradial motions enhance the typical gain region dwell time, we argue in this paper that this explanation for the ease of explosion in 2D and 3D relative to 1D is incomplete.} 
If nonradial motions, in the end, served only to increase the efficiency of neutrino heating beyond the critical threshold, then we should expect that critical cases in 1D, 2D, and in 3D reach a comparable thermal state at the onset of explosion, reflected in comparable postshock thermal pressure.
We present evidence on the basis of new high-resolution 1D, 2D, and 3D
core-collapse supernova simulations that this is not at all the case.
We show that, all else being equal, marginal explosions in 3D (and 2D)  reach explosion at much lower neutrino heating rates, far lower total absorbed energy, and at lower postshock thermal pressure than such explosions in 1D.
This indicates that some other phenomenon besides the deposition of neutrino energy is aiding explosion in 2D and 3D. 
An obvious candidate is turbulence in the gain layer.  
\cite{bhf:95} already pointed out that rising turbulent-convective bubbles excert an effective dynamical pressure on the shock and thus aid shock expansion, an argument recently reaffirmed by \cite{dolence:13} and \citet{couch:13b} on the basis of their 3D light bulb simulations. 
Importantly, Reynolds stress exerted by turbulence yields an effective ram pressure term in the Rankine-Hugoniot momentum condition describing the pressure balance between postshock and preshock pressure. In this way, turbulence can aid the postshock thermal pressure in pushing the shock out. 
\cite{murphy:13} were the first to point this out quantitatively.
More qualitatively, \citet{yamasaki:06} showed that inclusion of a convective flux in a steady-state accretion flow reduced the critical neutrino luminosity by about a factor of two.
This decrease is not due to a turbulent pressure, which is absent in the model of Yamasaki \& Yamada, but due to the effect of convective mixing increasing the entropy just behind the shock.

We present results from new 3D core-collapse supernova simulations of a $15$-$M_\odot$ (at zero-age main-sequence) progenitor of \cite{woosley:07} employing the FLASH code \citep{dubey:09, lee:14} and a multi-species neutrino leakage scheme that takes into account deleptonization and neutrino heating.
This scheme has been shown to well reproduce the global qualitative and quantitative features of postbounce core-collapse supernova evolution \citep{ott:13a,couch:14a}. It is superior to the simpler light bulb approach while still extremely computationally efficient, allowing extensive 3D parameter studies like
the one we present here. 
Following up on \cite{couch:13d}, we carry out a set of 3D core-collapse supernova simulations and vary the magnitude and spatial distribution of aspherical perturbations in the progenitor's Si/O shell. 
\emph{We show that for the wide range of our  3D simulations the strength of turbulence in the gain region is the key indicator of the susceptibility to explosion for a given simulated model}.   
We demonstrate that the turbulent ram pressure aiding explosion is systematically stronger in 2D simulations than in 3D simulations, which serves as a natural explanation for why neutrino-driven explosions obtain more easily in 2D than in 3D. 

Recently, \cite{nagakura:13} have proposed a new explosion criterion that they call ``critical fluctuation'' on the basis of semi-dynamical 1D simulations in which they give the stalled shock various positive radial velocities against the upstream accretion flow. 
They recorded the minimum shock velocity necessary to cause an explosion for fixed enclosed mass and shock radius.  
The relative postshock pressure change required to obtain this shock velocity was then defined as the ``critical fluctuation.'' We note that this critical fluctuation picture does not pertain to our results and conclusions. 
We are not considering a sudden increase of the postshock pressure, but the quasi-steady contribution of the turbulent ram pressure to the structure of the postshock region, which, as we show, aids neutrino-driven explosions in 2D and 3D.

In Section \ref{sec:method}, we describe our numerical approach and the simulations we have run.  
In Section \ref{sec:1Dv3D}, we point out that enhanced absorption of neutrino energy is insufficient to explain the relative ease of explosion in 2D and 3D simulations as compared with 1D simulations.  
We show that the turbulent ram pressure is systematically (and artificially!) higher in 2D than in 3D in Section~\ref{sec:2Dv3D}. 
We present our analysis of the strength of turbulence across our wide range of 3D simulations in Section \ref{sec:turb} and show that turbulence plays a central role in pushing 3D models over the critical threshold to explosion.  
We discuss our results and present our conclusions in Section \ref{sec:conclusions}.

\section{Simulations}
\label{sec:method}

\begin{figure*}[!htb]
  \centering
  \small\addtolength{\tabcolsep}{-3pt}
  \begin{tabular}{cccc}
    \includegraphics[width=1.77in]{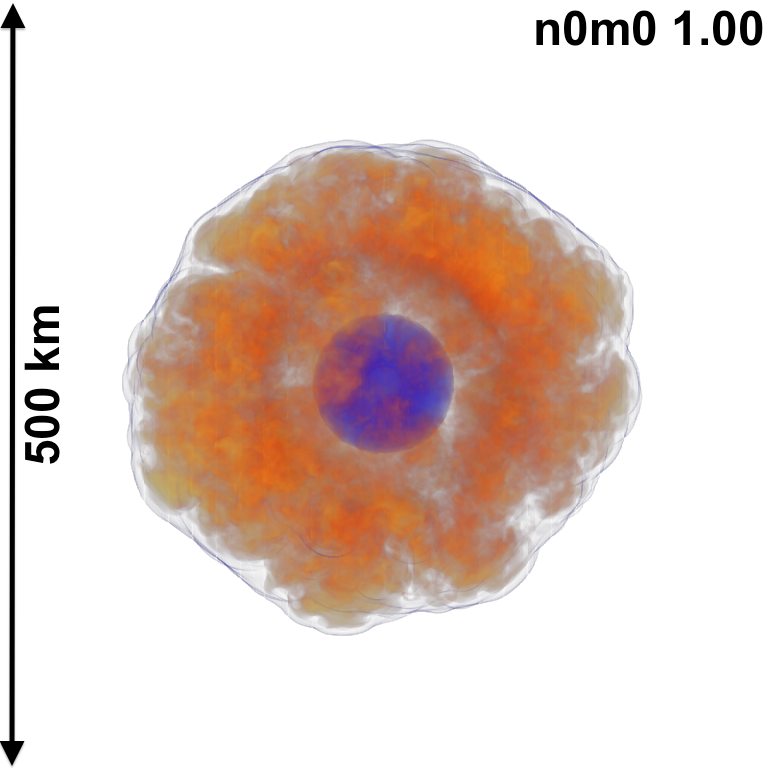} &
    \includegraphics[width=1.77in]{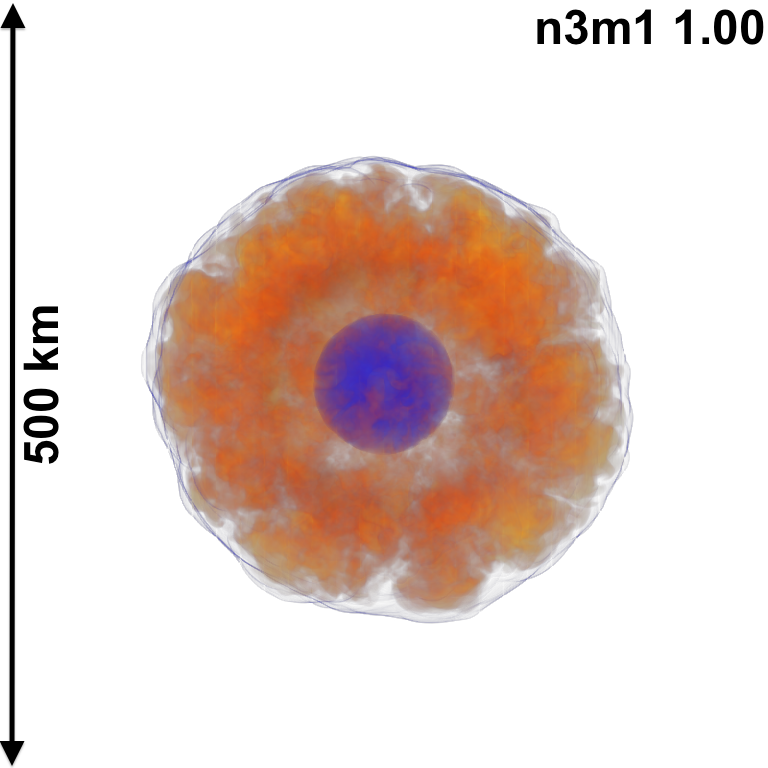} &
    \includegraphics[width=1.77in]{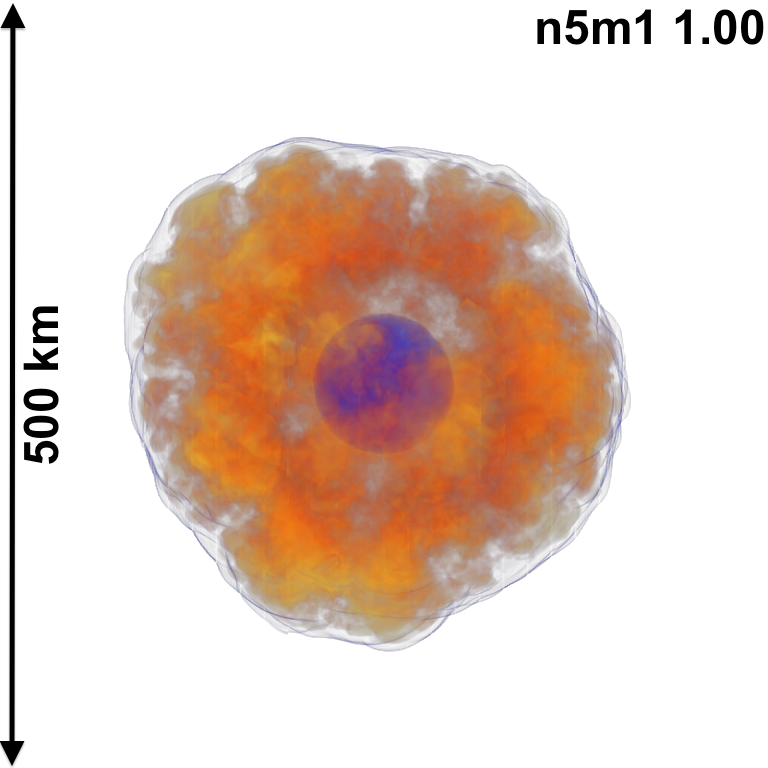} &
    \includegraphics[width=1.77in]{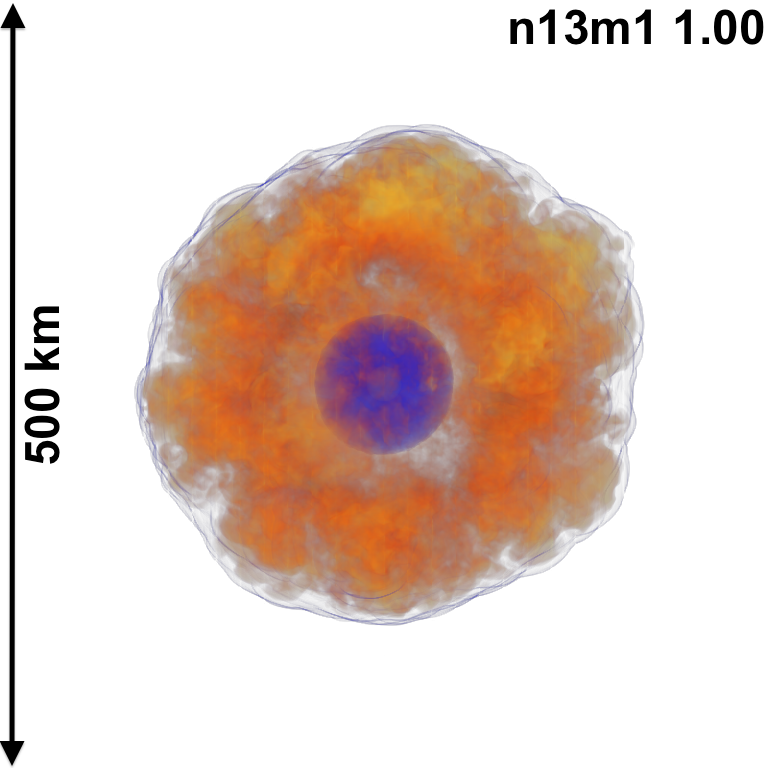} \\ 
    \includegraphics[width=1.77in]{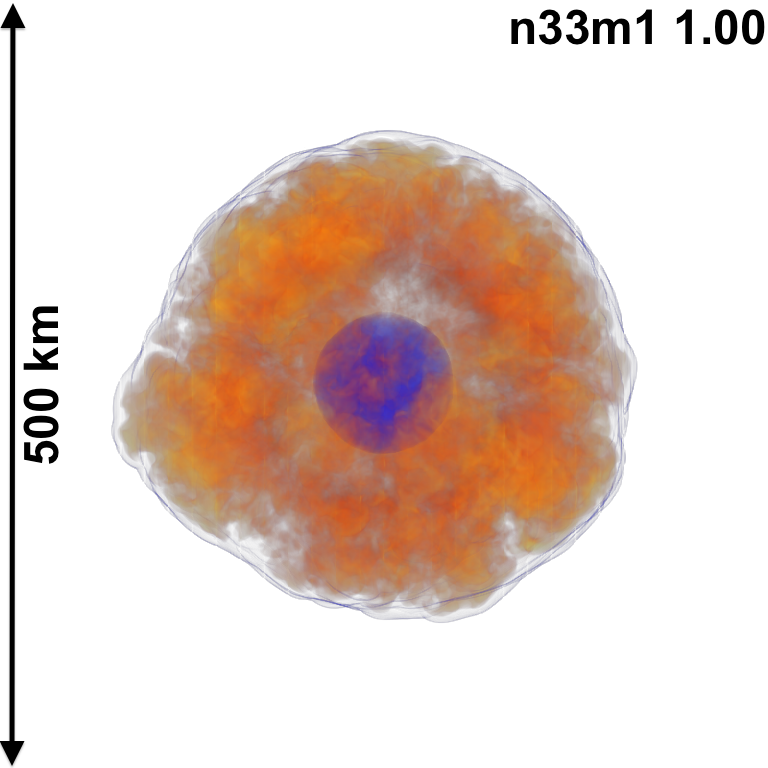} &
    \includegraphics[width=1.77in]{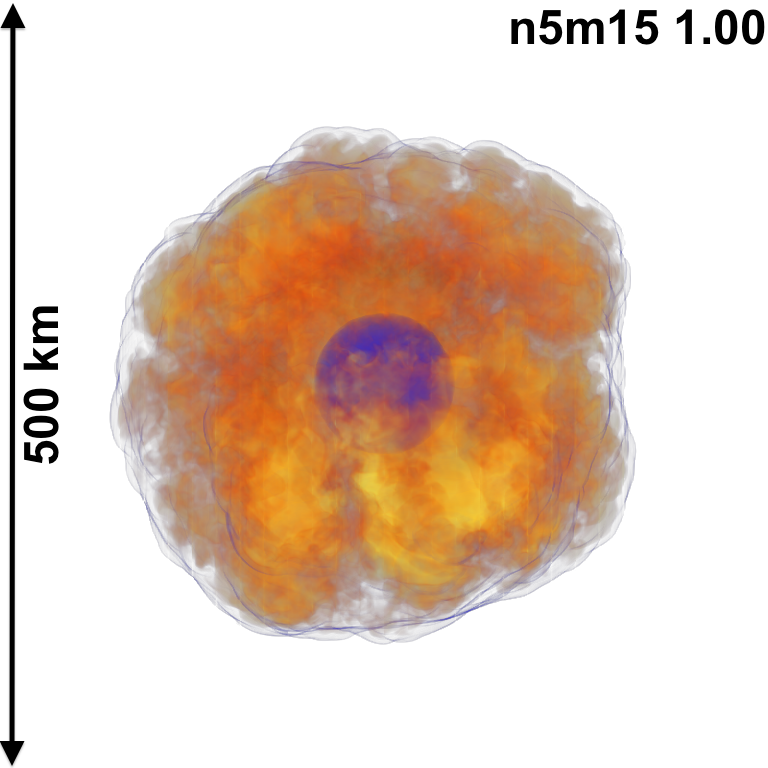} &
    \includegraphics[width=1.77in]{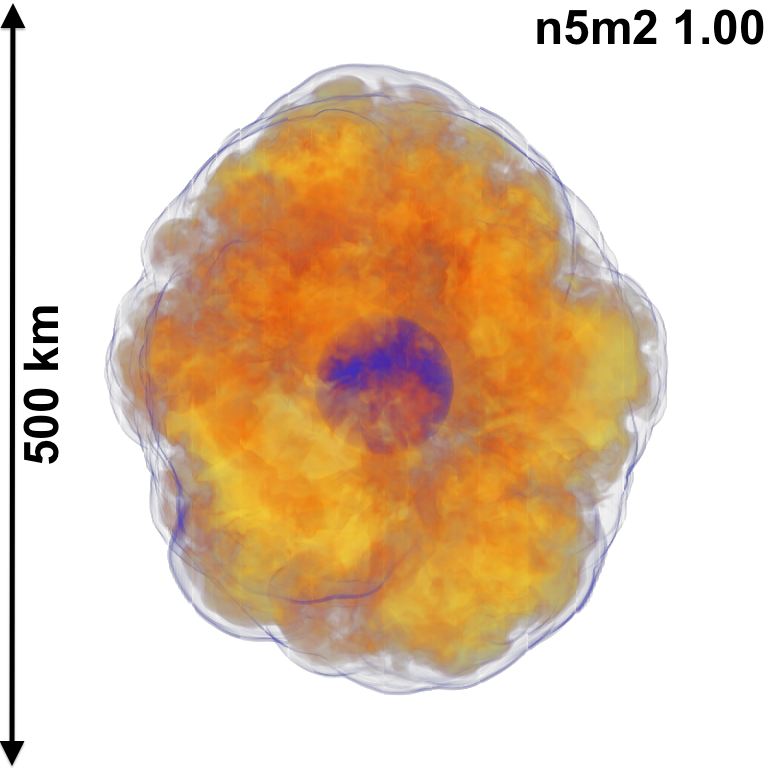} &
    \includegraphics[width=1.77in]{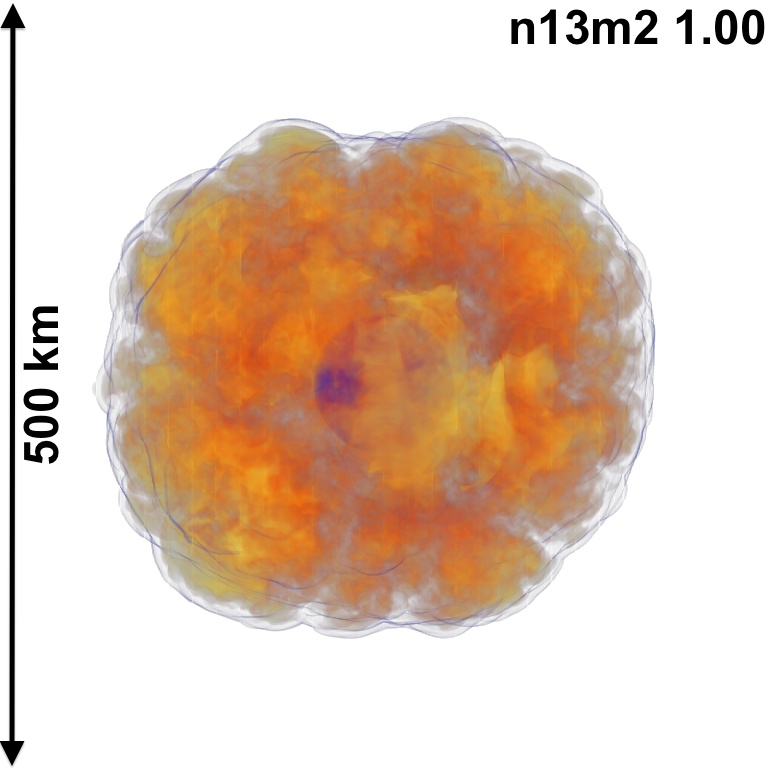} \\
    \includegraphics[width=1.77in]{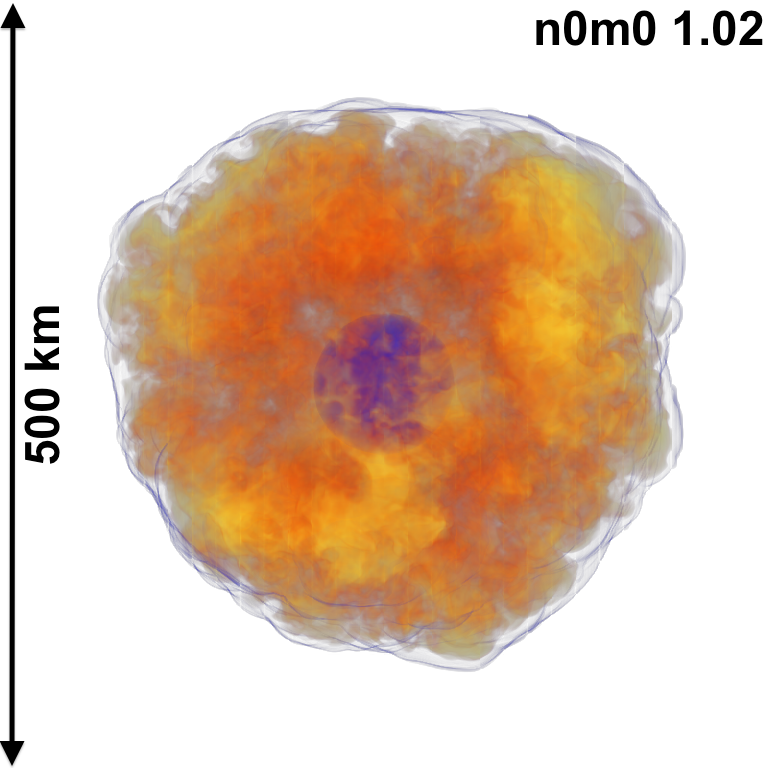} &
    \includegraphics[width=1.77in]{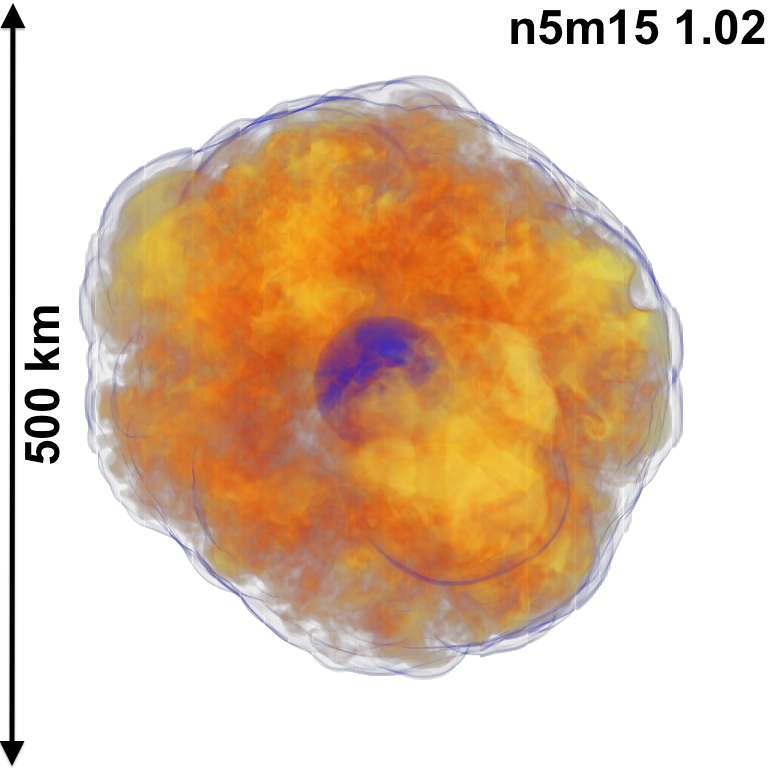} &
    \includegraphics[width=1.77in]{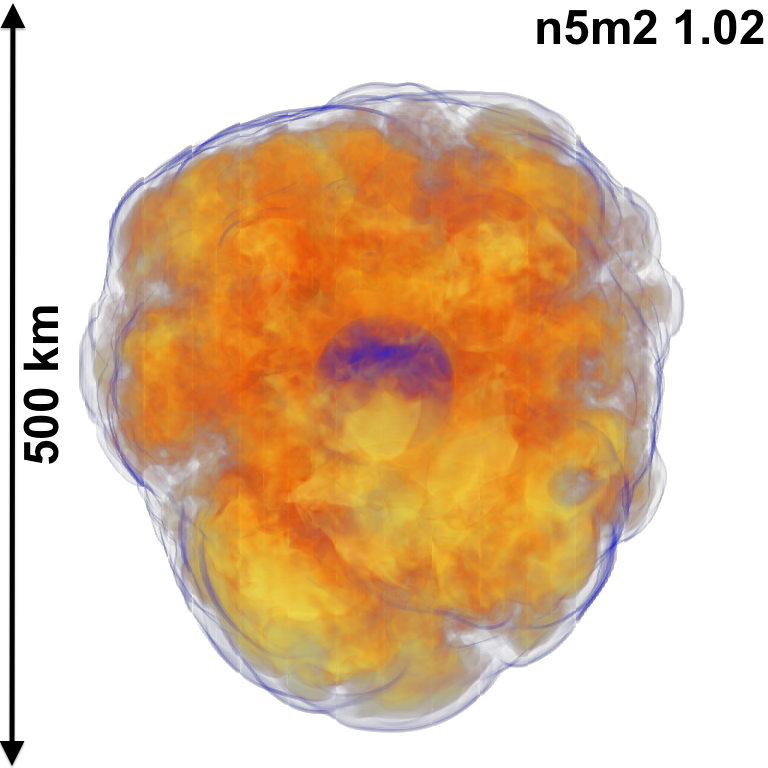} &
    \includegraphics[width=1.77in]{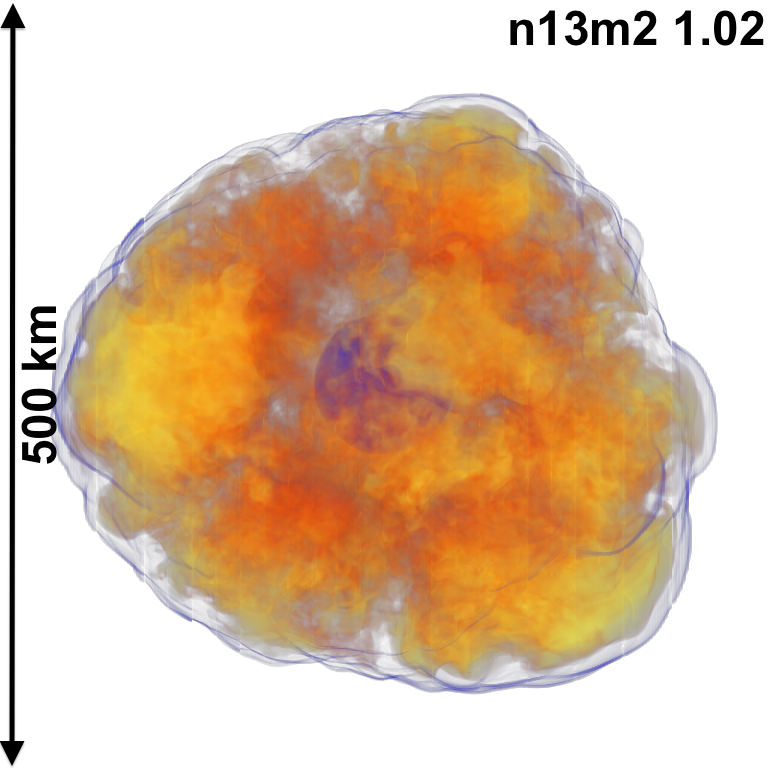}
  \end{tabular}
  \caption{
    Volume renderings of specific entropy for several of the 3D simulations at $150\,\mathrm{ms}$ after bounce.
    Darker, red colors correspond to specific entropies of $\sim$$14\,k_\mathrm{B}\,\mathrm{baryon}^{-1}$ while lighter, yellow colors correspond to entropies of $\sim$$18\,k_\mathrm{B}\,\mathrm{baryon}^{-1}$.
    The blue colors, which highlight the shock surface and the lower-entropy cooling region near the protoneutron star, correspond to specific entropies of $\sim$$5 \,k_\mathrm{B}\,\mathrm{baryon}^{-1}$.
    Models with stronger perturbations show higher specific entropies in the gain layer and a greater shock extension.
    This is a result of the stronger turbulence, and concomitant higher neutrino heating efficiency and turbulent pressure, in these models.
  }
  \label{fig:renderings}
\end{figure*}

\subsection{Methods}

Our numerical method is the same as that described in \citet{couch:13d} and \citet{couch:14a}.  
Our core-collapse supernova simulation application is built in the FLASH framework \citep{dubey:09, lee:14}.  
We utilize a directional-unsplit PPM hydrodynamics scheme \citep[c.f.,][]{colella:84} with an approximate HLLC Riemann solver in smooth flow and a HLLE Riemann solver near shocks.  
Self-gravity is included via a fast and accurate multipole approximation \citep{couch:13c} including Legendre orders up to 16.  
We employ adaptive mesh refinement (AMR) on a Cartesian grid with a minimum grid spacing of 0.49 km.  
The maximum allowed refinement level is limited as a function of radius from the origin, yielding a pseudo-logarithmic grid spacing with radius.  
The first forced decrement in refinement in our fiducial models occurs around 100 km, the second at 200 km, the third at 400 km, and so on.  
This gives an average effective resolution of $\alpha = \langle \Delta x_i / r \rangle \sim 0.75\%$, or $\sim 0\fdg43$, in the angular directions, where $\Delta x_i$ is the linear zone size and $r$ is the distance from the origin.

We include the effects of heating, cooling, and electron fraction evolution due to neutrinos via an approximate, ray-by-ray, multi-species leakage scheme \citep{oconnor:10,ott:13a}.  
This scheme accounts for the cooling due to heavy lepton neutrinos and antineutrinos and cooling and electron fraction changes due to electron neutrinos and anti-neutrinos. 
Charged-current heating is included via \citep{janka:01}, 
\begin{equation}
  Q_{\nu_i} = f_\mathrm{heat}\frac{L_{\nu_i}(r)}{4\pi r^2} \left \langle \frac{1}{F_{\nu_i}} \right \rangle \sigma_{\nu_i} \frac{\rho X_{(n/p)}}{m_\mathrm{amu}} e^{-2\tau_{\nu_i}},
 \label{eq:heating_rate}
\end{equation}
where $i$ runs over both electron neutrinos and anti-neutrinos, $L_{\nu_i}(r)$ are the neutrino luminosities as a function of radius, $r$, $\left \langle 1/F_{\nu_i} \right \rangle $ is the energy-averaged mean inverse flux factor, $\sigma_{\nu_i}$ are the charged-current cross sections, $\rho$ is the matter density, $X_{(n/p)}$ are the neutron/proton mass fractions, and $\tau_{\nu_i}$ are the neutrino optical depths.  
See \citet{couch:14a} for detailed definitions of these quantities and a complete description of how they are calculated in our leakage scheme.  
Equation (\ref{eq:heating_rate}) is a solution to the spherically symmetric neutrino transport equation, save for the addition of the \fheat\ factor.
This is included as a means of increasing (or decreasing) the local (i.e.\ microscopic) efficiency of the neutrino heating {\it without} changing other details of the neutrino radiation as would result from, e.g., a simple change in the opacities.  
The nominal value of \fheat\ is unity.  
Due to the strong non-linearities of the core-collapse supernova problem, small changes in the {\it local} heating efficiency, \fheat, can result in large changes in the {\it global} heating efficiency \citep[e.g.,][]{ott:13a}.

All simulations described here make use of the 15-\msun\ progenitor of \citet{woosley:07}.  
Such a star, if it were to occur in Nature, would be likely to result in a successful explosion and the formation of a neutron star based on both its compactness \citep{oconnor:11, ugliano:12, kochanek:14a} and its low helium core mass \citep{clausen:14a}.  
As in \citet{couch:13d}, we apply non-radial, momentum-preserving velocity perturbations to the otherwise spherically-symmetric initial model.  
These perturbations are applied only to the $\theta$-direction velocity component and are described by,
\begin{equation}
  \delta v_\theta = \mathcal{M}_{\rm pert} c_S \sin [(n-1)\theta] \sin [(n-1) \zeta] \cos (n \phi)\,,
  \label{eq:perts}
\end{equation}
where $\mathcal{M}_{\rm pert}$ is the peak perturbation Mach number, $c_S$ is the local adiabatic sound speed, $n$ is the number of nodes in the interval $\theta=[0,\pi]$, and $\zeta = \pi (r - r_{\rm
  pert,min})/(r_{\rm pert,max} - r_{\rm pert,min})$.
The perturbations are applied within the Si/O shell (roughly $r\sim 1000\ {\rm km} - 5000\ {\rm km}$), the inner iron core is left completely spherical.

\subsection{Overview of Models}

\begin{deluxetable}{clccccc}
\tablecolumns{7}
\tabletypesize{\scriptsize}
\tablecaption{
Overview of simulation parameters and results. 
\label{table:sims}
}
\tablewidth{0pt}
\tablehead{
\colhead{$n$\tablenotemark{a}} &
\colhead{$\mathcal{M}_{\rm pert}$\tablenotemark{b}} &
\colhead{$f_{\textrm{heat}}$\tablenotemark{c}} &
\colhead{$\alpha$\ [\%]\tablenotemark{d}} &
\colhead{$t_{\textrm{end}}$\ [ms]\tablenotemark{e}} &
\colhead{$r_{\textrm{ sh, max}}$\ [km]\tablenotemark{f}} &
\colhead{$\bar{\eta}_{\textrm{heat}}$\tablenotemark{g}}
}
\startdata 
\cutinhead{1D}
0  & 0    & 1.36 & 0.75 & 676 & 2008  & 0.169 \\
\cutinhead{2D}
0  & 0    & 1.05 & 0.75 & 367 & 1121  & 0.133 \\
\cutinhead{3D}
0  & 0    & 1.00 & 0.75 & 157 & 170.4 & 0.082 \\
0  & 0    & 1.00 & 1.40 & 362 & 175.6 & 0.083 \\
0  & 0    & 1.02 & 0.75 & 304 & 183.8 & 0.090 \\
0  & 0    & 1.05 & 0.75 & 273 & 390.9 & 0.101 \\
0  & 0    & 1.05 & 1.40 & 257 & 427.4 & 0.101 \\
3  & 0.1  & 1.00 & 0.75 & 210 & 167.4 & 0.080 \\
5  & 0.1  & 1.00 & 0.75 & 210 & 171.5 & 0.082 \\
5  & 0.15 & 1.00 & 0.75 & 271 & 176.5 & 0.083 \\
5  & 0.15 & 1.02 & 0.75 & 201 & 192.0 & 0.092 \\
5  & 0.2  & 1.00 & 0.75 & 307 & 180.7 & 0.083 \\
5  & 0.2  & 1.02 & 0.75 & 323 & 430.0 & 0.092 \\
5  & 0.2  & 1.02 & 1.40 & 332 & 452.4 & 0.094 \\
9  & 0.2  & 1.02 & 0.75 & 171 & 195.3 & 0.092 \\
13 & 0.1  & 1.00 & 0.75 & 210 & 172.4 & 0.083 \\
13 & 0.15 & 1.00 & 0.75 & 311 & 168.9 & 0.081 \\
13 & 0.2  & 1.00 & 0.75 & 303 & 180.4 & 0.084 \\
13 & 0.2  & 1.02 & 0.75 & 203 & 192.4 & 0.093 \\
17 & 0.1  & 1.00 & 0.75 & 175 & 172.0 & 0.082 \\
33 & 0.1  & 1.00 & 0.75 & 211 & 169.6 & 0.082
\enddata
\tablenotetext{a}{Perturbation scale (see Equation (\ref{eq:perts})).}
\tablenotetext{b}{Peak perturbation Mach number.}
\tablenotetext{c}{Local neutrino heating efficiency (see Equation (\ref{eq:heating_rate})).}
\tablenotetext{d}{Average resolution as a percent of spherical radius.}
\tablenotetext{e}{Time simulated after bounce.}
\tablenotetext{f}{Maximum shock radius attained during the simulated time.}
\tablenotetext{g}{Neutrino heating efficiency averaged between 50 and 150 ms after bounce.}
\end{deluxetable}

We run a total of 19 high-resolution, 3D core-collapse supernova simulations in which we vary the scale and amplitude of the initial perturbations, \fheat, and the resolution.\footnote{Some data for unperturbed models are taken from \citet{couch:14a}.}.
We refer to the simulations using the scheme n[{\it node count}]m[{\it initial perturbation Mach number, times ten}] [{\it heat factor}].
We consider 7 node counts: 0, 3, 5, 9, 13, 17, 33, and 4 amplitudes: 0 $\mathcal{M}$, 0.1 $\mathcal{M}$, 0.15 $\mathcal{M}$, and 0.2 $\mathcal{M}$.
These values yield velocity perturbations in the Si/O shell not unlike the fluctuations found in realistic, multi-dimensional simulations of convective nuclear burning in massive stars in both scale and strength \citep{meakin:07b, arnett:11a}.
The perturbations we apply also result in total added kinetic energies very similar to the convective kinetic energies found in the above-cited works:  $2.2\times 10^{47}$ erg, $4.9\times10^{47}$ erg, and $8.7\times10^{47}$ erg for $\mathcal{M}$= 0.1, 0.15, and 0.2, respectively.
These energies are orders of magnitude below both the thermal and gravitational binding energies of the progenitor's Si/O shell.
Most simulations were carried out with \fheat\ = 1 but for a few combinations of node count and amplitude we have increased \fheat.

In Table \ref{table:sims}, we list all the simulations along with key parameters and results.  
Only model n5m2 1.02, at both high- and low-resolution, explodes within the simulated time.  
For the sake of the present discussion, we shall consider a model to have ``exploded'' if the shock radius exceeds $400\,\mathrm{km}$ with no sign of returning.
This is, of course, only an approximate definition of a successful explosion. 
The other perturbed models with \fheat\ $>$ 1.00 (i.e., n9m2, n13m2) are near the explosion threshold, but as we show below, an analysis of the strength of turbulence in these models implies that conditions are less favorable for runaway shock expansion than in model n5m2.  
In Figure \ref{fig:renderings}, we show entropy volume renderings from 12 of our 19 3D simulations.
These visualizations illustrate that turbulence is stronger, and the shock radii larger, for models with stronger precollapse perturbations and/or higher \fheat.

\section{Explosions Aided by Turbulence}
\label{sec:1Dv3D}

\subsection{Comparison with a 1D Explosion}

\begin{figure}
  \centering
  \includegraphics[width=3.4in,trim= 0.08in 0.15in 0.3in 0.4in,clip]{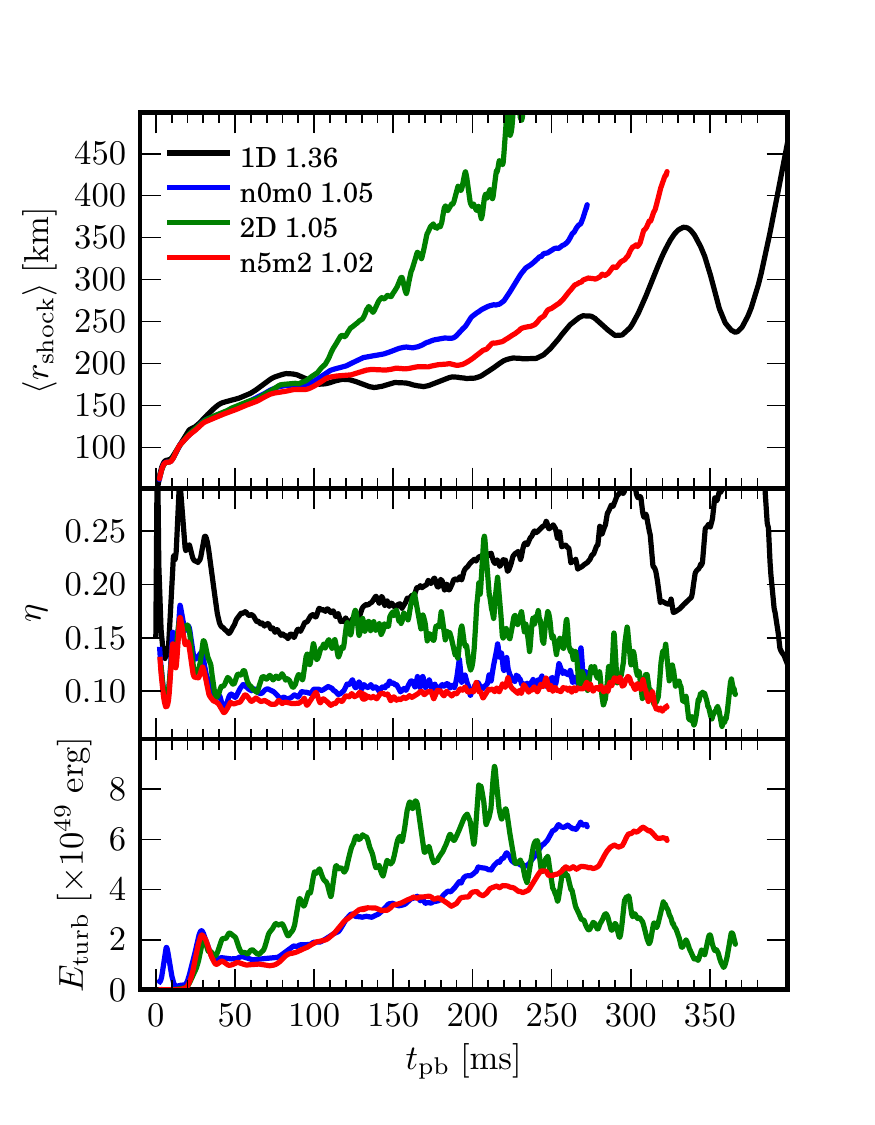} 
  \caption{
    Key diagnostics as functions of postbounce time for a critical 1D explosion (black lines), an unperturbed 3D explosion (blue lines), a perturbed 3D explosion (red lines), and an unperturbed 2D explosion.
    The 2D explosion model uses the same \fheat\ as the 3D unperturbed explosion (\fheat\ = 1.05). The top panel shows the angle-averaged shock radius as a function of postbounce time. The center panel depicts the global heating efficiency $\eta = Q_\mathrm{net} (L_{\nu_e} + L_{\bar{\nu}_e})^{-1}$, where $Q_\mathrm{net}$ is the integrated net heating in the gain layer and $L_{\nu_e}$ and $L_{\bar{\nu}_e}$ are the electron neutrino and anti-electron neutrino luminosities entering from below, respectively. The bottom panel shows the turbulent energy in 2D and 3D simulations defined by Equation~(\ref{eq:eturb}).
The 1D critical explosion shown here actually completes one more oscillatory cycle before the shock runs away. 
}
  \label{fig:1Dv3D}
\end{figure}

In Figure \ref{fig:1Dv3D}, we present diagnostic measures for
four comparable core-collapse supernova explosions in the 15-\msun\ progenitor: a 1D critical explosion (i.e.\ a simulation with the lowest $f_\mathrm{heat}$ leading to an explosion; black lines), a 2D explosion with \fheat\ = 1.05 (green lines), a 3D explosion with \fheat\ = 1.05 (blue lines), and the exploding perturbed 3D model of \citet{couch:13d} with \fheat\ = 1.02.
Save for the 2D case, each of these models results in a very marginal explosion.
The top panel of Figure~\ref{fig:1Dv3D} shows the average shock radius and the center panel the global neutrino heating efficiency, $\eta = Q_{\mathrm{net}} (L_{\nu_e} + L_{\bar{\nu}_e})^{-1}$, where $Q_{\mathrm{net}}$ is the net neutrino heating in the gain layer.
Immediately evident from the center panel is that the neutrino heating efficieny  required to trigger an explosion in 1D is far greater than in both 2D and 3D.
By the time the shock has reached $400\,\mathrm{km}$ (top panel of Figure~\ref{fig:1Dv3D}), at around $400\,\mathrm{ms}$ after bounce, the
1D model has absorbed  $1.1\times10^{52}\,\mathrm{erg}$  of neutrino energy whereas the 3D model n5m2 1.02 absorbs only $4.4\times10^{51}\,\mathrm{erg}$ prior to its average shock radius exceeding $400\,\mathrm{km}$, less than half that of the 1D case!

Evidently, critical explosions in 3D are not absorbing nearly as much neutrino energy as critical 1D explosions.
Nevertheless, the average shock radius in model n5m2 1.02 secularly expands at a slightly more rapid pace than the 1D model. 
What, then, is allowing the 3D neutrino mechanism to push the shock outward at a lower critical heating efficiency and with less total absorbed neutrino energy? 
We suggest that the answer is turbulence in the gain layer. 

As has been pointed out many times in the literature (see Section~\ref{sec:intro}), the nonradial motions inherent to turbulent convection allow a given fluid parcel to stay longer in the gain layer, so less of the absorbed energy is advected per unit time out of the gain layer into the cooling region. 
This is certainly an important part of the explanation.  
However, what has been generally ignored in the literature is that turbulence itself provides an effective pressure that adds to the thermal pressure behind the shock. The turbulent pressure thus allows the shock in multi-dimensional core-collapse supernovae to overcome the ram pressure of accretion at a \emph{lower} thermal
pressure and, hence, with less neutrino heating.

\begin{figure}
  \centering
  \includegraphics[width=3.4in,trim= 0in 0.1in 0.in 0.in,clip]{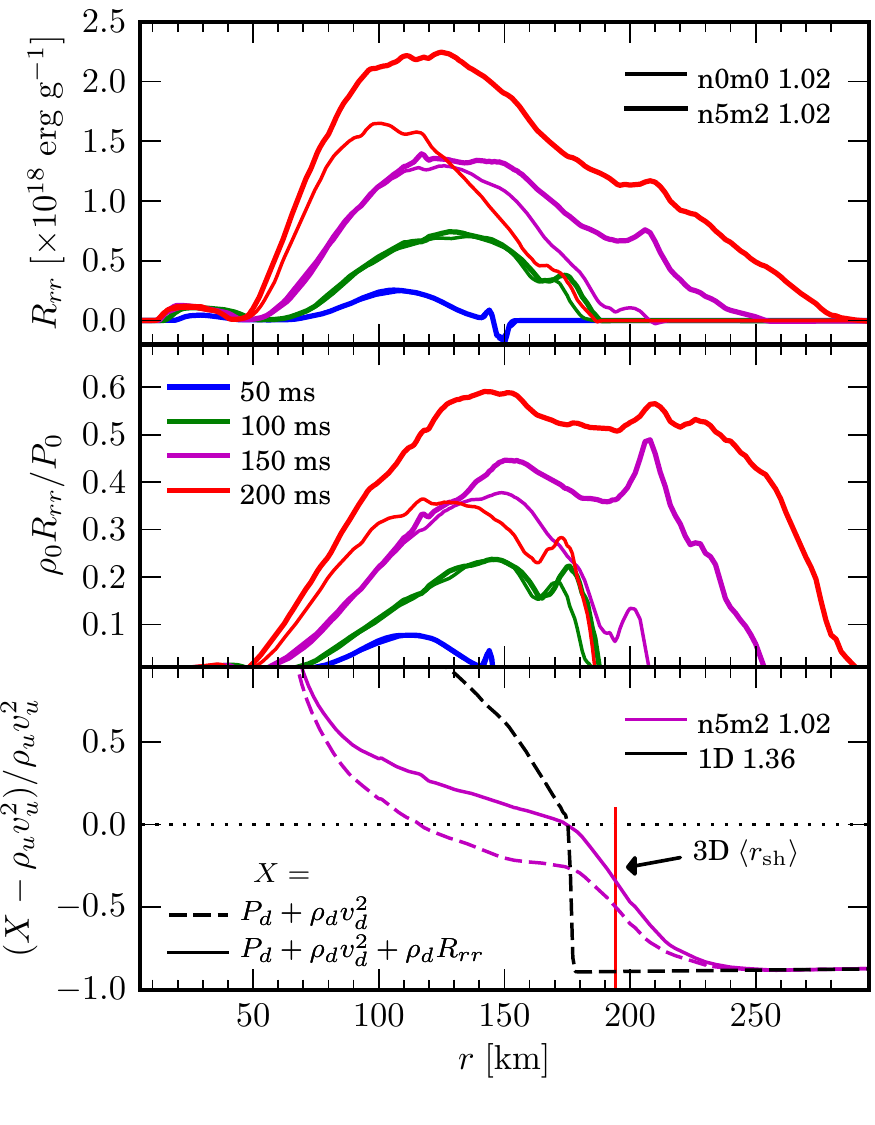} 
  \caption{
    Comparison of the $rr$-component of the Reynolds stress (Equation \ref{eq:stress}) between two 3D simulations: n5m2 1.02, which explodes, and n0m0 1.02, which fails to explode (top panel).
    Following accretion of the perturbed Si/O interface, the Reynolds stress in model n5m2 1.02 becomes increasingly larger than that in model n0m0 1.02.
    The center panel compares the implied turbulent ram pressure, $\rho_0 R_{rr}$, between the two 3D models, normalized to the background thermal pressure, $P_0$.
    The turbulent ram pressure can be a very significant fraction of the thermal pressure, up to 50\% in model n5m2 1.02.
    The bottom panel illustrates the impact of this turbulent ram pressure by comparing the residuals of the shock jump condition (Equation (\ref{eq:jump})) between a critical 1D explosion (\fheat~=~1.36, black line) and model n5m2 1.02 (red lines) at 150 ms.
    The shock location should be where the residual is zero, i.e., when the LHS and RHS of Equation (\ref{eq:jump}) balance one another, modulo effects due to asphericity of the shock.
    For the 3D case, we consider two scenarios: one in which the turbulent ram pressure is neglected (dashed lines) and the other in which it is included.
    Without the turbulent ram pressure, the shock radius is grossly underestimated.
    The average shock radius at this time for n5m2 1.02 is marked by the vertical magenta line.
  }
  \label{fig:reynolds}
\end{figure}

\cite{murphy:13} were the first to point out the importance of turbulent pressure in the steady-state conditions determining the position of the  stalled shock. Their Reynolds decomposition of the shock jump conditions implies that the effective downstream ram pressure includes two components: the background component and a turbulent component. 
The Reynolds-decomposed momentum condition reads
\begin{equation}
P_d + \rho_d v_d^2 + \rho_d R_{rr} \approx \rho_u v_u^2\,\,,
\label{eq:jump}
\end{equation}
where the right-hand side (upstream, preshock) term $\rho_u v_u^2$ is the ram pressure of accretion. The  left-hand side terms correspond to the downstream (postshock) flow and are the thermal pressure $P_d$, the ram pressure of the average background flow $\rho_d v_d^2$, and the turbulent ram pressure $\rho_d R_{rr}$. The latter is the $rr$-component of the Reynolds stress tensor.

The Reynolds stress tensor is $R_{ij} = v^\prime_i v^\prime_j$, but following \citet{murphy:13} we approximate the stress tensor by
\begin{equation}
  R_{ij} = \frac{\langle \rho u_i v_j^\prime \rangle}{\rho_0},
  \label{eq:stress}
\end{equation}
where $v_j^\prime = u_j - v_j$, $u_j = \langle v_j \rangle$, and $\rho_0 = \langle \rho \rangle$.
\citet{murphy:13} find that this expression is nearly identical to the standard stress tensor while better capturing the turbulent velocities near steep density gradients.
In the top two panels of Figure \ref{fig:reynolds}, we compare the two 3D simulations n0m0 1.02 and n5m2 1.02 in terms of the $rr$-component of the Reynolds stress (top panel) and turbulent ram pressure $\rho_0 R_{rr}$, which we normalize by the angle-averaged background thermal pressure $P_0 = \langle P \rangle $ (center panel). 
The perturbed model n5m2 1.02 (thick lines) develops an explosion, while the unperturbed model n0m0 1.02 (thin lines) fails to explode despite having identical local neutrino heating rate \fheat.  
In \citet{couch:13d}, we argued that this was due to the perturbations exciting stronger nonradial motion, enhancing the global neutrino heating efficiency by increasing the typical dwell time of matter in the gain layer.
Figure~\ref{fig:reynolds} shows that this explanation is incomplete.
Following the accretion of the perturbed Si/O interface, which occurs around $75\, \mathrm{ms}$ after bounce for this progenitor, the postshock Reynolds stress grows increasingly larger in model n5m2 1.02 than in model n0m0 1.02.
The center panel of Figure~\ref{fig:reynolds} shows that the turbulent ram pressure can be very significant.
Already by $100\,\mathrm{ms}$ after bounce, it is around 20\% of the background thermal pressure and grows to in excess of 50\% of the thermal pressure in model n5m2 1.02 by $200\,\mathrm{ms}$.
This additional ram pressure requires that the shock move outward, either to a new stability or to instability attended by runaway shock expansion.

The bottom panel of Figure~\ref{fig:reynolds} further illustrates this point (see also Figure~7 of \citealt{murphy:13}).
Here we show the residual between the LHS and RHS of the shock jump condition, Equation (\ref{eq:jump}), for the 1D critical explosion and for model n5m2 1.02.
The 3D data are first spherically-averaged and the upstream and downstream quantities are estimated by separate power-law fits to the data on either side of the smeared shock transition region.
The location where the upstream ram pressure of the accretion flow ($\rho_u v^2_u$) is balanced (i.e., where the residual is zero), should be the approximate shock location.
We have considered two scenarios: one in which the LHS of Equation (\ref{eq:jump}) includes the turbulent ram pressure, and the other in which it does not.
Of course, in 1D, there is no turbulent ram pressure and we find that the sum of the downstream thermal and ram pressures balances the ram pressure of the accretion flow precisely at the 1D shock radius ($\sim$$174$ km).
For the 3D model, however, leaving out the turbulent ram pressure results in a major underestimate of the shock location ($\sim$$116$ km).
Inclusion of the turbulent ram pressure in 3D pushes the shock radius as estimated from the shock jump condition out significantly, much nearer to the actual average shock radius for n5m2 1.02 at this time ($\sim$$194$ km).
This demonstrates that in 1D, the thermal pressure must do all the work in pushing the shock out whereas in 3D, the turbulent ram pressure makes a tremendous contribution to this effort.
The 3D simulation does not need to reach nearly the same postshock thermal pressures as the 1D critical case in order for the shock to run away.

An important point must be made concerning the turbulence-modified shock jump conditions, Equation (\ref{eq:jump}).
The turbulent ram pressure, $\rho_0 R_{rr}$, is defined only based upon angle averages of the underlying hydrodynamic quantities.
Thus, the modified jump conditions that result from a Reynolds decomposition of the momentum equation \citep{murphy:13} only apply to the angle-averaged conditions.
That is to say, for any given normal to the shock surface in 2D or 3D, the standard shock jump conditions, neglecting the turbulent pressure, apply by definition.
This is precisely what is required by our finite volume hydrodynamic method that solves Euler's equations in their standard form.
It is only when one spherically-averages the fundamentally multi-dimensional flows in 2D and 3D that the standard jump conditions, which only account for one-dimensional flow, breakdown and the inclusion of the turbulent effective pressure is needed.

\subsection{Comparison With a 2D Explosion}
\label{sec:2Dv3D}

\begin{figure}
  \centering
  \includegraphics[width=3.4in,trim= 0in 0.1in 0.in 0.in,clip]{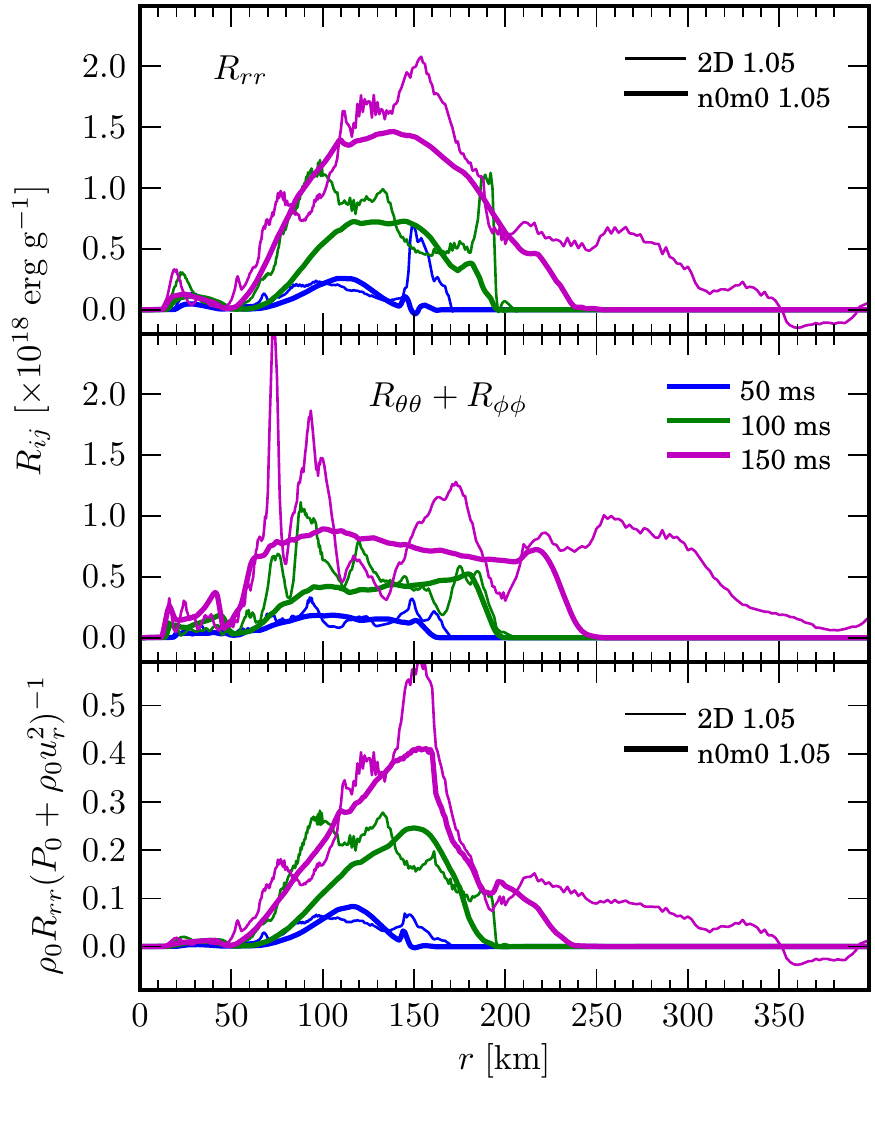} 
  \caption{
    Comparison of the Reynolds stresses between 2D (thin lines) and 3D (thick lines).
    The top panel shows the ${rr}$-component, the center panel shows the sum of the $\theta \theta$- and $\phi \phi$-components (for 2D, $R_{\phi \phi} = 0$), and the bottom panel shows the turbulent ram pressure normalized by the background total pressure $P_0+\rho_0 u_r^2$, where $\rho_0 u_r^2$ is the angle-averaged radial ram pressure.
    At every time shown, the 2D Reynolds stress, and the corresponding turbulent ram pressure, is greater than in the comparable 3D simulation.
    This is a natural result of the unphysical behavior of turbulence in 2D (i.e., the inverse energy cascade) and is a principal cause of the greater ease of explosion in 2D as compared with 3D.    
  }
  \label{fig:reynolds2d}
\end{figure}

Figure \ref{fig:1Dv3D} also shows the average shock radius (top panel), global heating efficiency (center panel), and turbulent energy in the gain layer (bottom panel) for an exploding 2D model with \fheat~=~1.05.
We define the total turbulent kinetic energy as
\begin{equation}
  E_{\mathrm{turb}} = \int_{\mathrm{gain}} e_{\mathrm{turb}}\ dV\,\, ,
  \label{eq:eturb}
\end{equation}
where the turbulent energy density is
\begin{equation}
  e_{\mathrm{turb}} = \onehalf \rho [(v_r - \langle v_r \rangle )^2 + v_\theta^2 + v_\phi^2]\,\,,
  \label{eq:eturbDens}
\end{equation}
and $\langle ... \rangle$ represents an angle average.
Notable is that already at early times, $\sim$$50 - 100\,\mathrm{ms}$ after bounce, while the average shock radii are still very similar, the 2D simulation has higher total turbulent energy than either of the 3D simulations shown in Figure~\ref{fig:1Dv3D}.
And, despite identical heat factor \fheat\ to the 3D model n0m0 1.05, this 2D simulation explodes much sooner.
Indeed, the lowest, or ``critical,'' \fheat\ at which this $15$-$M_\odot$ progenitor will explode in 2D is 0.95 \citep{couch:14a}, whereas the critical \fheat\ in 3D is 1.05 if the initial model is treated as spherically symmetric.
Many arguments have been made as to why 2D should be {\it more} favorable to explosion than 3D \citep[see][]{hanke:12, couch:13b, couch:14a, takiwaki:14a}, but a consideration of the strength of the turbulence adds one more.
\citet{hanke:12} and \citet{couch:14a} have already shown that the erroneous inverse turbulent energy cascade in 2D pumps kinetic energy to large scales in 2D.
These works argued, largely from an empirical basis, that turbulent energy at large scales was more favorable to explosion.
In light of the arguments in the preceding section and in \cite{murphy:13}, we consider and compare the Reynolds stresses and turbulent ram pressures in 2D and 3D.

The top two panels of Figure~\ref{fig:reynolds2d} show the Reynolds stresses in the 3D model n0m0 1.05 compared to the stresses in the 2D simulation with \fheat~=~1.05 at $50$, $100$, and $150\,\mathrm{ms}$ after bounce. 
At these times, the 2D and 3D shock radii have not yet dramatically diverged,  permitting a direct comparison.
At every epoch, the Reynolds stresses are greater in 2D than in 3D, in agreement with the higher total turbulent energy we find in 2D (cf.\ bottom panel of Figure~\ref{fig:1Dv3D}).
The bottom panel of Figure~\ref{fig:reynolds2d} shows the turbulent ram pressure normalized to the background thermal pressure.
The turbulent ram pressure in 2D is much greater than in 3D,
in particular right behind the stalled shock at $50\,\mathrm{ms}$ and
$100\,\mathrm{ms}$ after bounce.
At these times the angle-averaged thermal pressure profiles between these 2D and 3D simulations are not very different at all.
This, in combination with the higher turbulent energy in 2D, is another physical explanation for why explosions are more easily obtained in 2D than 3D.
As clearly demonstrated in other works, the erroneous inverse turbulent energy cascade in 2D results in far greater turbulent energy on large scales in 2D as compared to 3D \citep{hanke:12, couch:13b, couch:14a}. 
This indicates a connection between the postshock turbulent ram pressures and large-scale turbulent motions.

\subsection{Justification for the Presence of Turbulence}

In the previous sections, we argued that increased neutrino heating efficiency due to multi-dimensional effects is not the only cause of the greater ease of explosion in 2D and 3D as compared with 1D.
We showed that critical explosions in 3D occur at much lower global heating efficiency and, for the $15$-$M_\odot$ progenitor considered here, absorb {\it less than half} of the radiated neutrino energy than critical 1D explosions over a comparable period of postbounce time.
Of course, 1D lacks any possibility of nonradial dynamics that could increase the dwell time in the gain layer. Hence, a greater fraction of the absorbed energy will inadvertently be advected out into the cooling region, making the total heating requirements for explosion naturally more stringent than in 1D than in 2D or 3D. However, our analysis of the turbulent stresses behind the stalled shock in 2D and 3D shows that the turbulent ram pressure amounts to a significant fraction of the total pressure during the stalled shock phase and that this is crucial in pushing the shock outward in 2D and 3D, an effect that is absent in 1D.
We now turn to the question of whether or not our simulations are plausibly turbulent, in the numerical sense.

Turbulence has been most thoroughly investigated under conditions satisfying three common criteria: steady-state, isotropy, and incompressibilty \citep[e.g.,][]{kolmogorov:41, pope:00}.
It is highly questionable that any one of these criteria is satisfied in the gain layer of a core-collapse supernova, let alone all three.
The background accretion flow and time-dependent accretion rate disrupt the steady state.
The strong gradients and buoyancy-driven directionality of the convection make turbulence anisotropic.
And the turbulent Mach numbers can exceed 0.5 \citep{couch:13d}, making the flow at least somewhat compressible.
Lacking a more general theory of turbulence, we shall, nevertheless, charge ahead and consider the character of turbulence in our simulations under the standard assumptions given above.

The strength of turbulence is typically characterized by the dimensionless Reynolds number, the ratio of inertial forces to viscous forces: $Re = v_0 l_0 / \nu$, where $v_0$ is a typical turbulent speed, $l_0$ is the inertial length scale, and $\nu$ is the kinematic viscosity.
In the protoneutron star, below the neutrinosphere, where neutrinos are fully diffusive, they dominate the viscosity \citep{kazanas:78, burrows:88b, thompson:05}. 
However, as the neutrinos decouple from matter, their mean
free path exceeds the scale of the gain layer, effectively removing them
as a source of viscosity. 
Estimates of the physical Reynolds numbers in the gain layer can be made on the basis of the Braginskii-Spitzer shear viscosity for a collisional
plasma \citep{braginskii:58,spitzer:62,braginskii:65}. In the core-collapse supernova context, momentum transport is dominated by proton collisions in the gain layer, and the Braginskii-Spitzer viscosity yields very large Reynolds numbers (i.e., $\gg 1000$).  
The effective, {\it numerical} Reynolds number will be much lower owing to the numerical viscosity exceeding the actual physical viscosity.
The Reynolds number range in which the postshock flow in the core-collapse supernova context transitions from laminar to turbulent is completely unknown and intuition regarding this transition gleaned from other turbulent contexts may not apply here.
Furthermore, we caution the reader that the Reynolds number may not even be the correct dimensionless ratio of forces to consider, since it neglects the influence of buoyancy, advection, and neutrino heating which play a crucial role in the gain region turbulence.

Estimating the numerical Reynolds number accurately is difficult since it requires a measurement of the rate of numerical dissipation in the specific problem being considered.
Making the assumptions of incompressibility, isotropy, and steady-state, the Reynolds number can be expressed in terms of the Taylor microscale $\lambda$ as \citep[D. Radice, private communication;][]{abdikamalov:14b}
\begin{equation}
Re = 5 \left (\frac{l_0}{\lambda}  \right)^2\,\,.
\label{eq:reynolds}
\end{equation}
The Taylor microscale is given by $\lambda^2 = 5 E / Z$ \citep{pope:00}, where the total turbulent energy and enstrophy are, respectively,
\begin{equation}
E = \int_0^\infty E_{\textrm{turb}}(\ell) d \ell = \onehalf \rho_0 v_0^2\,\,,
\end{equation}
and
\begin{equation}
Z = \int_0^\infty \ell^2 E_{\textrm{turb}}(\ell) d \ell\,\,,
\end{equation}
where $E_{\textrm{turb}}(\ell)$ is the turbulent kinetic energy density in spherical harmonic space \citep[see, e.g.,][]{couch:14a}.

\begin{figure}
  \centering
  \includegraphics[width=3.4in]{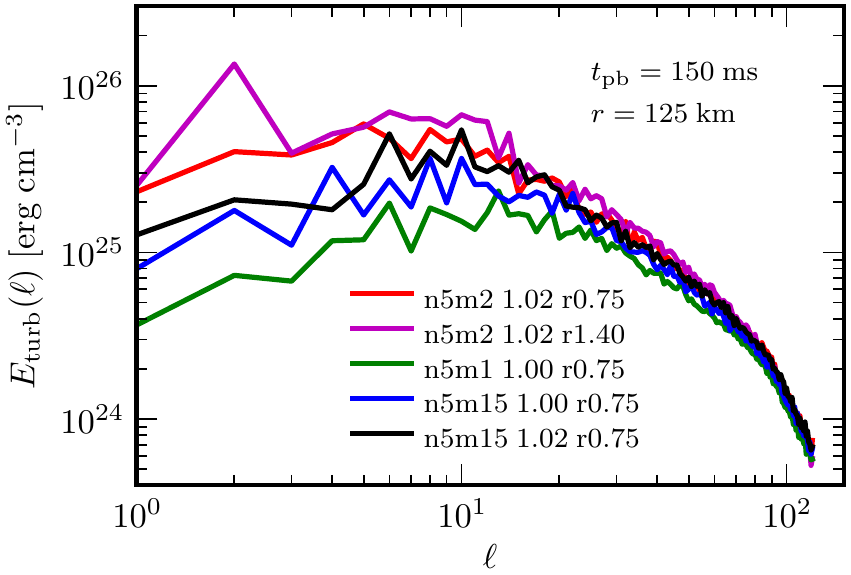}
  \caption{
    Turbulent energy spectra in the gain layer at $150\,\mathrm{ms}$ after bounce in several 3D models with perturbations of the same spatial scale ($n =5$; see Equation \ref{eq:perts}), but different peak perturbation Mach number $\mathcal{M}$ and different \fheat.
We also plot the reduced-resolution simulation n5m2 1.02~r1.40. 
This simulation has twice the linear zone spacing in the gain twice as the fiducial r0.75 simulations. Simulations perturbed with $\mathcal{M}=0.2$ (corresponding to m2 in the model name) show the highest turbulent energy. Note that the reduced-resolution simulations has significantly more energy at large scales and more energy overall than the fiducial-resolution simulations.
  }
  \label{fig:spec_n5}
\end{figure}

$E$ and $Z$ can be computed from the turbulent energy spectra.  Figure
\ref{fig:spec_n5} shows the turbulent energy spectra for model n5m2
1.02 at two different average effective angular resolutions, $\langle
\Delta x_i / r \rangle \sim 0.75\%$ and $\langle \Delta x_i / r
\rangle \sim 1.40\%$, where $\Delta x_i$ is the linear zone size and
$r$ is the distance from the origin (see Section~\ref{sec:method}).
The turbulent energy spectra are computed by taking the spherical
harmonic transform of the turbulent energy density
(Equation~\ref{eq:eturbDens}) and averaging the data over a
$10\,\mathrm{km}$-wide shell centered at a radius of
$125\,\mathrm{km}$ and averaging in time over $10\,\mathrm{ms}$ around
$150\,\mathrm{ms}$ after bounce.  Based on the high-resolution
spectrum, we estimate a Taylor scale of $12\,\mathrm{km}$.  We find
that the associated Legendre polynomials necessary for computing the
spherical harmonic transform can exceed IEEE double precision limits
for $\ell > 150$.  Hence, we truncate the spectra at $\ell \approx
150$ and potentially overestimate $\lambda$, since $\ell = 150$ is far
larger than the grid scale of our simulations: in the gain region,
$\ell_{\rm max} \sim \pi (125\ {\rm km} / 1\ {\rm km}) \sim 400$.  The
inertial scale $l_0$ should be the scale at which the energy spectrum
peaks.  In Figure \ref{fig:spec_n5}, we see that the kinetic energy
spectrum of model n5m2 1.02 r0.75 (red curve) peaks at $\ell \sim 5$.
The spectra are computed at a radius of 125 km, so the linear inertial
scale is $l_0 \sim \pi/5 \times 125\ {\rm km} \sim 80\ {\rm km}$.
This results in an estimated lower limit on $Re$ of $\sim 220$.  This
estimate could be too low by as much as a factor of a few due to the
truncation of the spherical harmonic transform.  Another estimate for
the Reynolds number can be made based on a ratio of the inertial scale
to the dissipation scale \citep{pope:00}, $Re \approx (l_0 /
l_D)^{4/3}$.  If we naively identify the
dissipation scale as the grid resolution in our simulations at 125 km,
where $\Delta x_i = 1\,\mathrm{km}$, we find $Re \sim (80\ {\rm km} /
1\ {\rm km})^{4/3} \sim 350$.
The results of
  \cite{benzi:08} suggest that the true effective dissipation scale
  for the PPM algorithm may be closer to roughly half a cell width, which
  would result in a larger estimated Reynolds number of $\sim$1000.
We note, however, that independent of the dissipation scale, numerical
viscosity has been shown to affect the flow and thus the turbulent
cascade up to scales of $\mathcal{O}(10)$ computational cell widths in
PPM simulations (e.g., \citealt{porter:98,sytine:00}).


Our rough estimate of $Re \sim 200 - 300$ for our 3D simulations based on Equation (\ref{eq:reynolds}) is well below  the Reynolds numbers typically considered turbulent \citep[e.g.,][]{pope:00}, but the transition from laminar to turbulent flow is not a sharp one and it is unclear, given the breaking of the standard assumptions, how to interpret the Reynolds number in the core-collapse supernova context.
A detailed 3D resolution study of turbulence in core-collapse supernova might elucidate this, though this is beyond the scope of the present work.
However, we have carried out one lower-resolution simulation of model n5m2 1.02 using an effective average angular resolution of $\langle \Delta x_i / r \rangle \sim 1.40\%$.
This model has about half the resolution in the gain region as the fiducial model.
Many of the integral metrics between the high- and low-resolution cases are similar, but the low resolution case explodes earlier, exhibits greater turbulent energy at large scales (Figure~\ref{fig:spec_n5}), and has greater total turbulent energy.
This is the expected behavior of turbulence in this instance.
The lowered resolution dissipates and transports energy to small scales {\it less} efficiently than the higher resolution case, leaving more energy at large scales and more turbulent energy overall.
The physical reason for this behavior is easily understood:  In the core-collapse supernova context, the primary mechanism that transports turbulent kinetic energy to small scales is the growth of parasitic instabilities on rising buoyant plumes in the gain layer. 
In the inviscid limit, these instabilities, mainly Rayleigh-Taylor and Kelvin-Helmholtz, grow ever faster with increasing grid resolution.
Faster growth of these instabilities leads to a more efficient turbulent energy cascade to small scales.
This is akin to arguments regarding resolution dependence given in \citet{couch:13b} and \citet{couch:14a} and underlines the need for a thorough resolution study of turbulence in the core-collapse supernova gain layer.

\begin{figure*}[!htb]
  \centering
  \begin{tabular}{cc}
    \includegraphics[width=3.5in,trim= 0in 0.15in 0.25in 0.4in,clip]{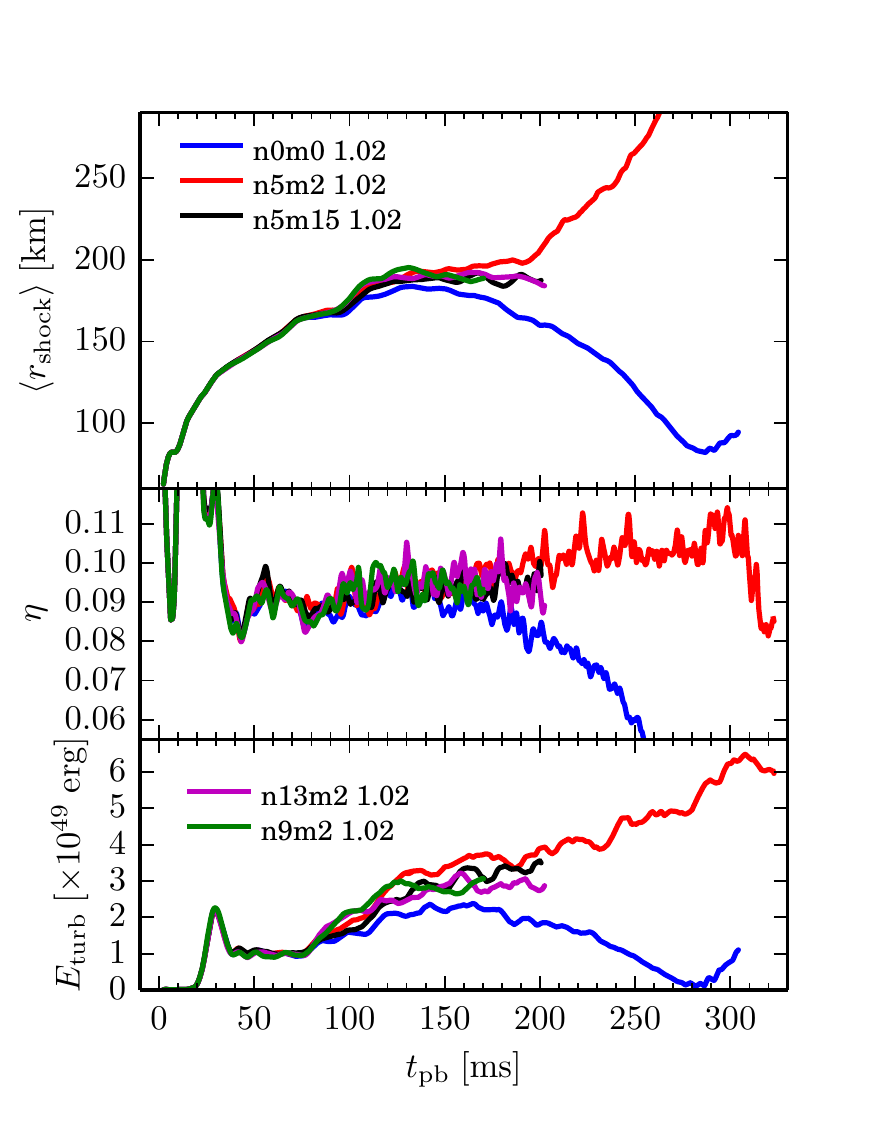} 
    \includegraphics[width=3.5in,trim= 0in 0.15in 0.25in 0.4in,clip]{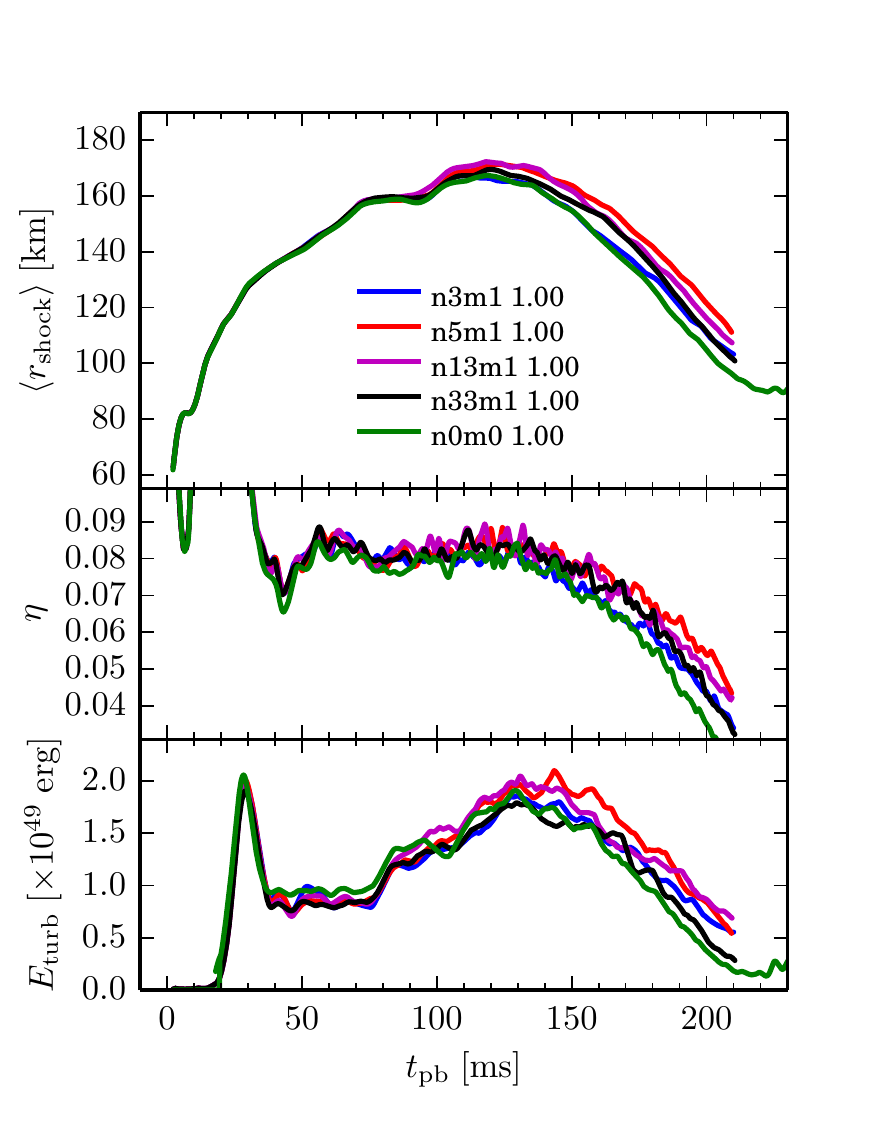} 
  \end{tabular}
  \caption{
    Average shock radius (top), neutrino heating efficiency (center), and total turbulent energy (bottom) as a function of postbounce time for several of our 3D simulations.
    The left panel shows results of simulations with \fheat~=~1.02, including the unperturbed simulation n0m0.
    The right panel shows simulations with low-amplitude perturbations and \fheat~=~1.00. Any level of perturbation increases the turbulent energy, but the effect is strongest for the highest perturbation Mach number $M=0.2$ (m2 in the model name).
  }
  \label{fig:rshock_all}
\end{figure*}

Regarding the requirements for grid resolution, we note that it should not be necessary to resolve the true, physical dissipation scale to obtain converged results for the core-collapse supernova problem.
In implicit large-eddy simulations (ILES), such as ours, accurately capturing the turbulent dissipation should be secondary to resolving the inertial range of the turbulence and capturing the large scales on which energy is transported.
In the particular context of the present discussion, this is tantamount to requiring convergence of the {\it turbulent ram pressure}, not the dissipation.
The turbulent ram pressure in our simulations should be most sensitive to the energy-carrying scales since $\rho_0 R_{rr}$ is basically an energy density, and in Kolmogorov's theory of turbulence (and in ILES) the energy is transported by the largest scales \citep{grinstein:07}.


\section{Dependence of Postshock Turbulence on Progenitor Asphericity}
\label{sec:turb}

\begin{figure*}[!htb]
  \centering
  \begin{tabular}{cc}
    \includegraphics[width=3.5in, trim= 0.05in 0in 0.3in 0.1in, clip]{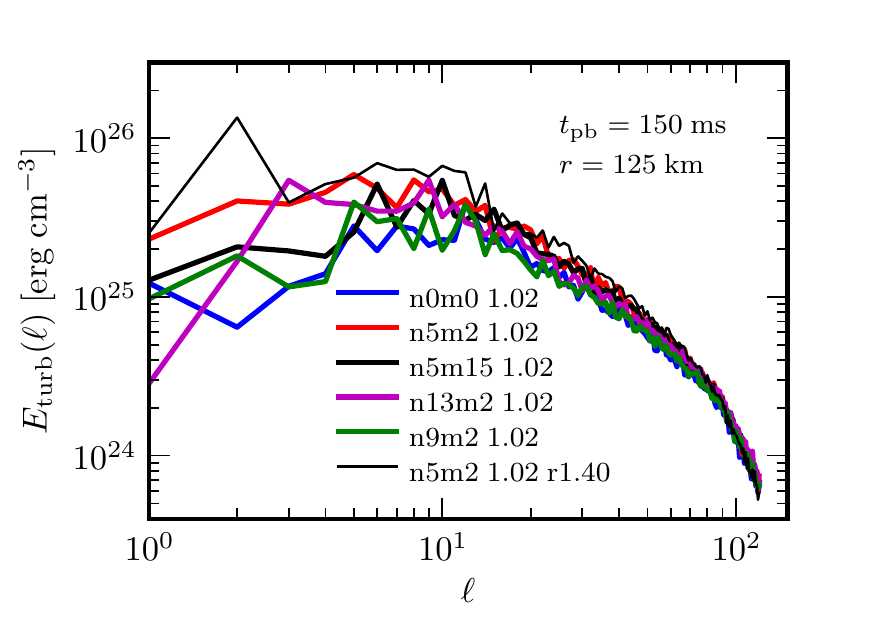} &
    \includegraphics[width=3.5in, trim= 0.05in 0in 0.3in 0.1in, clip]{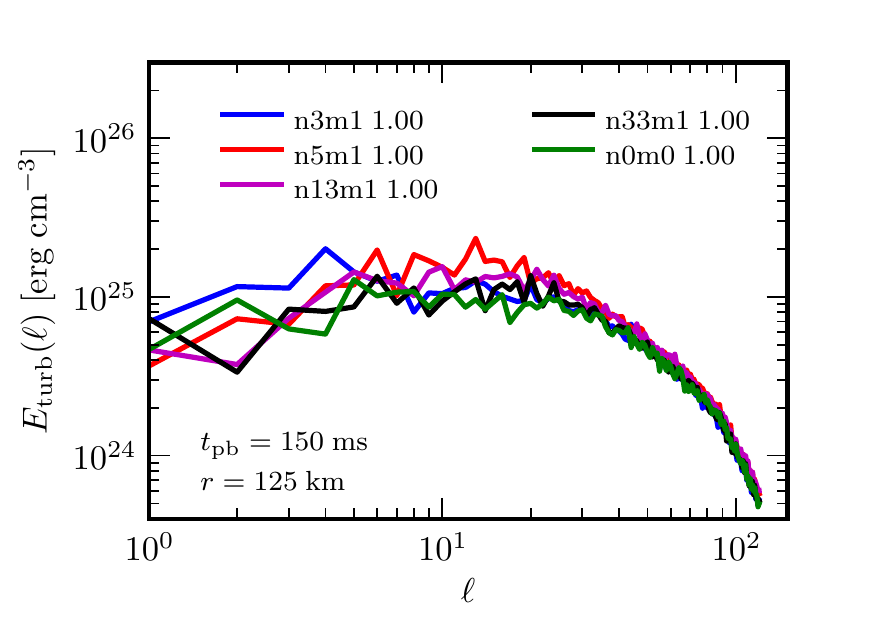}
  \end{tabular}
  \caption{
    Turbulent kinetic energy spectra in the gain layer at $150\,\mathrm{ms}$ after bounce for 3D simulations with \fheat~=~1.02 (left panel) and 3D simulations with small amplitude perturbations and \fheat~=~1.00 (right panel). Note the (weak) trend that larger-scale (lower n) perturbations lead to more turbulent energy at large scales.
  }
  \label{fig:spec_all}
\end{figure*}

\begin{figure}[!htb]
  \includegraphics[width=3.4in,trim= 0.07in 0.15in 0.33in 0.4in,clip]{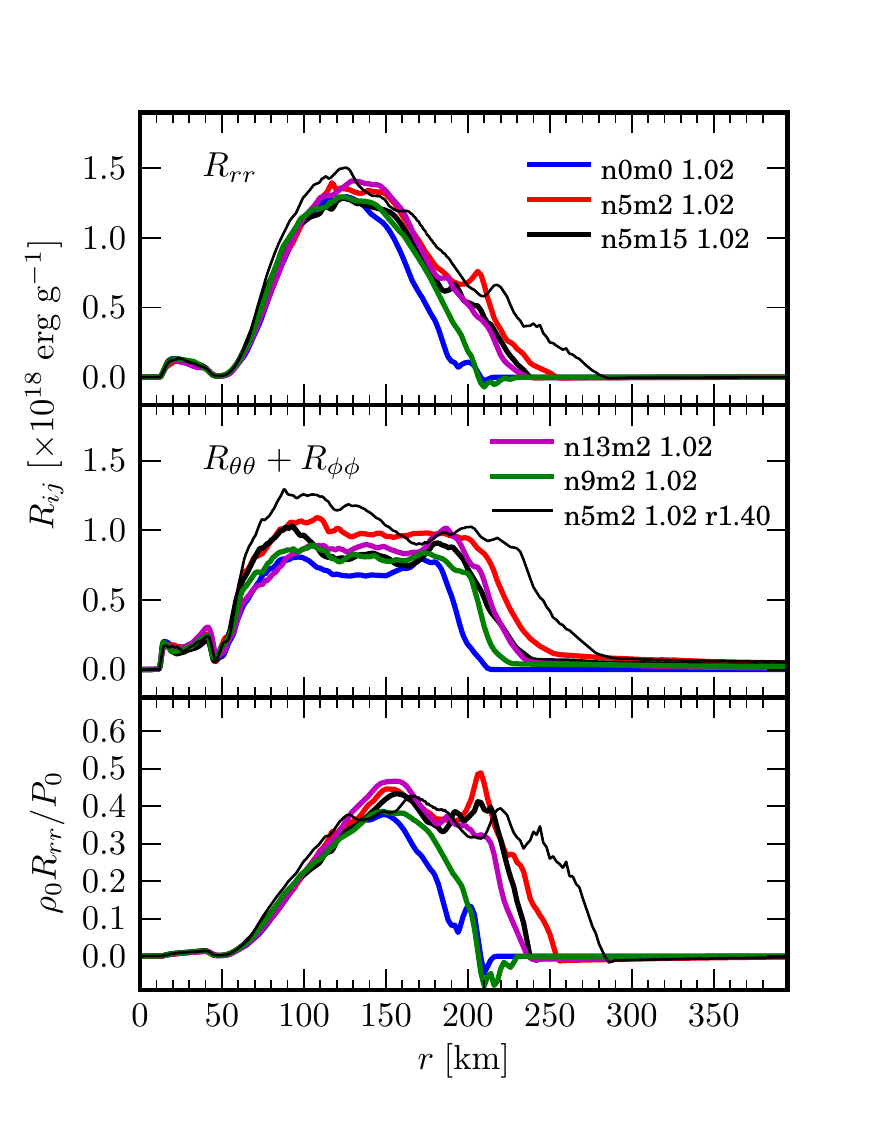} 
  \caption{
    Reynolds stresses at $150\,\mathrm{ms}$ after bounce in a representative subset of our 3D simulations, including the reduced-resolution simulation n5m2 1.02~r1.40.
    The top panel shows the $rr$-component, the center panel the sum of the $\theta \theta$- and $\phi \phi$-components, and the bottom panel shows the resulting turbulent ram pressure $\rho_0 R_{rr}$, normalized by the background thermal pressure $P_0$. Note that the reduced-resolution simulation has notably higher Reynolds stresses and turbulent ram pressure than its fiducial-resolution n5m2 1.02 counterpart.
  }
  \label{fig:reynolds3d}
\end{figure}

We have shown in the above that critical explosions in 2D and 3D are launched with much less neutrino heating than in 1D.
We argued that the key to understanding the reason for this is that turbulence provides an effective ram pressure that helps push the shock out against the accretion flow of the collapsing outer core in 2D and 3D.
Stronger turbulence results in a greater turbulent ram pressure and, hence, more favorable conditions for explosion.
In this section, we show that a plausible and effective means of enhancing the strength of turbulence in the gain region is the inclusion of physically-motivated progenitor asphericity, corroborating our initial findings that we reported in \cite{couch:13d}.

In Figure~\ref{fig:rshock_all}, we plot integral metrics for a large sample of our 3D simulation set.
On the left, we show the average shock radius, neutrino heating efficiency, and total turbulent energy for simulations with \fheat~=~1.02 and velocity perturbations with amplitudes $\mathcal{M} = 0.15$ and $0.2$ and, for reference, the unperturbed simulation with $\mathcal{M} = 0$.
On the right, we show the same, but for \fheat~=~1.00, weaker perturbations with $\mathcal{M} = 0.1$, and again the reference simulations with $\mathcal{M} = 0$.
For both \fheat, the inclusion of progenitor asphericity, however strong, results in increased turbulent energy and more favorable conditions for explosion (larger global heating efficiencies and greater peak shock radii).
The only model that explodes is n5m2 1.02, though all perturbed models with \fheat~=~1.02 shown in the left panels of Figure~\ref{fig:rshock_all} {\it might} explode if we ran them for a longer period of time.
Figure \ref{fig:rshock_all} shows, however, that model n5m2 1.02 results in a slightly larger average shock radius at $200\,\mathrm{ms}$ (when the other simulations end) and exhibits higher total turbulent energy than the other models from $125\,\mathrm{ms}$ after bounce onward.
This latter point, without doubt, contributes to the greater expansion of the shock in model n5m2~1.02 for reasons discussed in Section~\ref{sec:1Dv3D}.

In Figure \ref{fig:rshock_all}, we find a clear correlation between strength of turbulence and average shock radius.
We also find a weak trend between the strength of postshock turbulence and the scale of the applied perturbations:  larger scale perturbations (smaller $n$) result in stronger turbulence and larger average shock radii.
This weak correlation is also found in the turbulent kinetic energy spectra of these simulations, and it is also apparent in the entropy volume renderings in Figure \ref{fig:renderings}.
In Figure \ref{fig:spec_all}, we show the turbulent energy spectra for the same set of simulations shown in Figure \ref{fig:rshock_all}.
For both \fheat, the spectra tend to show more energy at large scales (small $\ell$) for low-order perturbations.
Additionally, stronger perturbations (higher $\mathcal{M}$) result in substantially greater turbulent energy at large scales and more total turbulent energy.
Interestingly, increasing the strength of the perturbations, and consequently the strength of turbulence, does not merely shift the spectra upward.
The additional turbulent energy seems concentrated at large scales, pushing the peak of the spectra to smaller $\ell$.
If we identify the peak in the spectra as the inertial scale, stronger perturbations then result in a larger inertial scale and greater inertial range.

The physical origin for the correlation between perturbation scale and strength of turbulence can be understood as follows.
The turbulence in our simulations is likely due to neutrino-driven convection \citep{murphy:13}.
Postshock turbulence can also be driven by the SASI \citep[e.g.,][]{endeve:12}, but the particular progenitor studied here
is less susceptible to the SASI than others (\citealt{couch:14a}; see also \citealt{fernandez:14}).
\citet{foglizzo:06} argued that the strength of postshock convection is increased by large preshock perturbations and this dependence was first demonstrated by \citet{scheck:08}.
Our results corroborate this: larger amplitude perturbations result in stronger nonradial motion behind the shock.
We also find that larger {\it scale} perturbations do the same.
As shown by \citet{foglizzo:06}, and embodied in their $\chi$ parameter, a perturbation entering the postshock region must be of sufficient spatial size such that it has time to grow to a scale at which it is buoyant before being advected out of the gain layer.

In Table \ref{table:sims}, we provide the average global neutrino heating efficiencies of all simulated models, time-averaged between $50$ and $150\,\mathrm{ms}$ after bounce. 
Over this time, the heating efficiencies are relatively constant.
The average heating efficiency depends most strongly on \fheat\ and varies only slightly with the strength of progenitor asphericity. 
Yet, when the Si/O interface is accreted through the shock around $75\,\mathrm{ms}$ after bounce, the average shock radii begin to diverge (cf.\ Figure~\ref{fig:rshock_all}).
This is due to the stronger turbulence in the gain region helping to push the shock outward.
This is shown graphically in Figure \ref{fig:reynolds3d} where we plot the diagonal components of the Reynolds stress tensor and the turbulent ram pressure (normalized by the background pressure) for 3D simulations with \fheat~=~1.02.
Immediately evident from Figure \ref{fig:reynolds3d} is that all the perturbed models show much larger Reynolds stresses, and attendent turbulent ram pressures, than the unperturbed model, n0m0 1.02.
While there is not a great deal of difference in the total turbulent kinetic energy between models n5m15 1.02, n9m2 1.02, and n13m2 at this time (cf.\ Figure \ref{fig:rshock_all}), the turbulent ram pressures are quite different, with n9m2 1.02 showing far less turbulent ram pressure behind the shock.
In fact, the average shock radius in this model is slowly receding, resulting in a negative $R_{rr}$ near the shock.
Model n9m2 1.02 is also distinguished in Figure \ref{fig:spec_all} by a deficit of turbulent kinetic energy at large scales, relative to the other perturbed models.
This is evidence that turbulent ram pressure is connected with turbulent kinetic energy on large scales.
As has been identified empirically in previous multi-dimensional simulations \citep{hanke:12, couch:13b}, turbulent energy on large scales is correlated with greater shock expansion and explosion.

Figure \ref{fig:reynolds3d} also shows the Reynolds stresses for the low-resolution model n5m2 1.02~r1.40.
Consistant with this model's more rapid shock expansion than the fiducial high-resolution case, n5m2 1.02 r1.40 has larger Reynolds stresses and postshock turbulent ram pressure.
This low-resolution model also shows substantially more turbulent energy on large scales (Figure~\ref{fig:spec_all}), adding further support to our claim that the turbulent ram pressure contribution behind the shock is most sensitive to the turbulent kinetic energy at large scales.

\section{Discussion and Conclusions}
\label{sec:conclusions}

Core-collapse supernovae are turbulent beasts. 
Unless the collapsing core of a massive star is extremely rapidly rotating or strongly magnetized (or both), there is no known physical process that would prevent  buoyancy and/or shear-driven turbulence in the gain layer right behind the stalled shock \citep{obergaulinger:14,fh:00,fryer:07}.

Our set of high-resolution 1D, 2D, and 3D core-collapse supernova simulations presented in this paper demonstrate that turbulence is an essential ingredient in neutrino-driven explosions and that there is a direct correlation between the strength of turbulence and the susceptability to explosion of a given multi-dimensional simulation. 

It has long been accepted that turbulence, in particular that arising from neutrino-driven buoyant convection, is beneficial to the neutrino mechanism.
Attempts to explain the favorable effect of turbulence have focused on the increase in gain-layer dwell time that nonradial motions in multi-dimensional simulations provide, allowing the material in the gain layer to more efficiently absorb neutrino energy. 
The quantitative argument supporting this explanation has often \citep[e.g.,][]{murphy:08,dolence:13,couch:13b} been that multi-dimensional simulations yield explosions at a lower ``critical'' driving neutrino luminosity $L_{\nu} = L_{\nu_e} + L_{\bar{\nu}_e}$ at a given pre-shock accretion rate $\dot M$ than 1D simulations. This is in the spirit of the argument of \cite{burrows:93} that states that no steady-state stalled-shock solutions exists above a critical $L_\nu - \dot M$ curve.

This, however, is only half the truth. 
Nature does not admit the freedom to dial-in neutrino luminosities. 
It is thus more physically enlightening to ask the question: How efficiently must the \emph{available} neutrino luminosity be deposited to drive an explosion and how does this efficiency depend on dimensionality of the simulation?  
Our results show that for the same progenitor, and thus the same available total luminosity, multi-dimensional simulations require much lower neutrino heating efficiencies (and much lower integrated absorbed energies) to launch explosions than their 1D counterparts. 
For the first time, we have demonstrated that this is not only because of a turbulence-mediated increase in the dwell time of matter in the gain region.
It is also the non-negligible contribution of the turbulent ram pressure that adds to the thermal pressure in the postshock region. The turbulent ram pressure \emph{allows multi-dimensional simulations to overcome the ram pressure of accretion at lower thermal background pressure} and thus with less neutrino heating. 

The results of our extensive parameter study, the largest to-date in 3D, encompassing a total of 19 3D simulations, demonstrate that precollapse asphericity from convective Si/O shell burning has a profound effect on the neutrino-driven turbulence in the gain layer. 
Our simulations show that larger-amplitude perturbations clearly lead to stronger turbulence and that there is also a trend, albeit weaker, linking larger-scale perturbations to stronger turbulence.
Stronger turbulence always brings a simulation closer to explosion.
We note that the turbulent kinetic energy in the gain region (lower panel, Figure \ref{fig:1Dv3D}) is substantially larger than the {\it total} kinetic energy of the added perturbations ($10^{47}$--$10^{48}$ erg), and during the simulated time only a fraction of this perturbation kinetic energy has been accreted through the shock.

In comparing turbulence in 2D and 3D, we find that 2D simulations artificially overpredict the strength of turbulence for the same progenitor and same local neutrino heating rate. 
Consequently, explosions are launched erroneously more easiliy in 2D than in 3D. 
Since turbulence appears to be at the heart of the core-collapse supernova problem and because Nature is 3D, the results of 2D simulations cannot be trusted and core-collapse supernovae must be studied in 3D.

The understanding of turbulence in the core-collapse supernova gain layer is still in its infancy, but our work and the work of others is beginning to emphasize its direct effects on core-collapse supernova dynamics, beyond its secondary effect of increasing the dwell time in the gain layer. 
The early work of \cite{bhf:95} deserves credit for pointing out the mechanical force that rising plumes exert on the stalled shock (albeit their picture was rather laminar-convective).  
\cite{murphy:13} for the first time revealed the role of turbulent ram pressure in setting the shock jump conditions and \cite{dolence:13}, as well as \cite{ott:13a} and \citet{couch:14a}, pointed out the dynamical impact of rising large, near-volume-filling turbulent bubbles on the stalled shock.  
Much analytic and computational work will be needed to derive a 3D model for core-collapse supernova turbulence, but the detailed 2D turbulence analysis of \cite{murphy:11} certainly marks an important stepping stone in this direction.

It is critical to point out that the significant role of turbulence in pushing the shock out in no way lessens the importance of neutrino heating to the core-collapse supernova mechanism.
Indeed, the energy to power the turbulence and explosions, as ever, comes from the neutrinos. In the simulations we have presented, the turbulence is driven by neutrino-driven convection \citep{murphy:13} and so is crucially dependent on the strength of the neutrino heating.
Strong turbulent convection, however, allows the core-collapse supernova mechanism to tap another power source:  the kinetic power of the accretion flow, which can amount to as much as $10^{52}\,\mathrm{erg}\,\mathrm{s}^{-1}$ following bounce.
We suspect that the shear between buoyant convective plumes and the accretion matter redirects a fraction of the kinetic energy of the accretion flow into turbulent kinetic energy in the gain layer, abetting or at least helping successful multi-dimensional explosions at lower neutrino heating rates than in 1D.
Of course, that is not to say that a stronger accretion flow would be {\it more} favorable to explosion; quite the opposite.
Also, both the accretion power and the neutrino luminosity draw their energy from the same well: the original gravitational binding energy of the progenitor core.

We have demonstrated that turbulence plays a {\it crucial} role in the core-collapse supernova mechanism.
This is a somewhat perilous conclusion to draw, however, because the turbulence in the gain region of our 3D simulations is without doubt not fully resolved.
We estimate numerical Reynolds numbers of $200 - 300$, where the physical Reynolds number are likely larger by orders of magnitude. 
We are certainly not capturing the turbulent dissipation rate accurately, however, within the ILES paradigm in which our simulations fall \citep{grinstein:07}, it is sufficient to resolve the inertial range.
These are the scales on which the turbulent energy is transported and the simulation results {\it should not} depend strongly on the dissipation rate.
Still, this assertion needs to be proven by a detailed and thorough resolution study of postshock turbulence in the core-collapse supernova context.
Our cursory investigation of resolution effects shows that decreased resolution results in greater turbulent energy on large scales and more favorable conditions for explosion.
This is the expected behavior of turbulence in this context: the lowered resolution results in a less efficient cascade of turbulent energy through the inertial range to small scales.
Similar resolution dependence was found by \citet{couch:13b}, \citet{couch:14a}, and in the high-resolution cases of \citet{hanke:12}.
\citet{handy:14} and the low-resolution cases of \citet{hanke:12} show the opposite dependence: lowering the resolution yields {\it less} favorable conditions for explosion.
Considering these disparate results from the point of view of turbulence may suggest a reconciling explanation.
The simulations of \citet{handy:14} and low-resolution cases of \citet{hanke:12} were carried out with very coarse resolution, only 2-3$^\circ$ in the angular directions.
This is far coarser than even the low-resolution case we consider here ($\sim 0\fdg86$).
It could be that their low-resolution simulations are not numerically turbulent and, hence, the turbulent ram pressures are too low to support shock expansion.
As the resolution is increased under these conditions, the turbulent ram pressure would become greater, pushing the shock out more, resulting in more favorable conditions for explosion.
At some critical resolution (we speculate around 1$^\circ$), the simulations would be effectively turbulent, but the intertial range, and hence forward energy cascade, would not be fully resolved.
Under {\it these} conditions, increasing the resolution would increase the efficiency of turbulent energy transport to small scales yielding less favorable conditions for explosions, since the turbulent ram pressure is most sensitive to large scale turbulent motions.
In other words, for sufficiently low resolution, the postshock fluid behaves like molasses in Winter, rather than like water because the numerical viscosity is just too large.
Once the inertial range is resolved, the simulation results should converge, even as the dissipation rate continues to change with increased resolution.
The scale at which the inertial range in the core-collapse supernova context is ``resolved'' is still an open question and the subject of future work.

The core-collapse supernova mechanism manifestly works in nature. Simulations must provide an answer to the core-collapse supernova problem even if the explosion mechanism is sensitive to stochastic variations in progenitor structure \citep{couch:13d,clausen:14a}. 
Turbulence, driven by neutrino energy deposition and enhanced by precollapse asphericity in the Si/O shell (which is essentially unavoidable), may be the answer. 
The results of our high-resolution 3D simulations certainly suggest so, but suffer in generality and reliability from our approximate treatment of neutrinos and our neglect of general relativity. 
Future fully self-consistent neutrino radiation-hydrodynamics simulations and extensive resolution studies will be needed to further ascertain the role of turbulence in the core-collapse supernova mechanism.

\section*{Acknowledgements}
The authors thank David Radice and Ernazar Abdikamalov for help with the computation of the numerical Reynolds number. 
The authors acknowledge further helpful discussions with W.~David Arnett, Adam Burrows, Evan O'Connor, Uschi C.~T.\ Gamma, Carlo Graziani, Philipp M\"osta, Christian Reisswig, Luke Roberts, and Petros Tzeferacos. 
SMC is supported by NASA through Hubble Fellowship grant No.\ 51286.01 awarded by the Space Telescope Science Institute and by NSF grant no.\ AST-0909132.  
CDO is partially supported by NSF grant nos.\ AST-1212170, PHY-1151197, and OCI-0905046, by a grant from the Institute of Geophysics, Planetary Physics, and Signatures at Los Alamos National Laboratory, by the Sherman Fairchild Foundation, and by the Alfred~P.~Sloan Foundation. 
The software used in this work was in part developed by the DOE NNSA-ASC OASCR Flash Center at the University of Chicago.  
The simulations were carried out on computational resources at ALCF at ANL, which is supported by the Office of Science of the US Department of Energy under Contract No.\ DE-AC02-06CH11357, on the NSF XSEDE network under computer time allocation TG-PHY100033, and on the NCSA Blue Waters supercomputer under NSF PRAC grant no.\ ACI-1440083.

\bibliography{bibliography/bh_formation_references,bibliography/sn_theory_references,bibliography/sn_observation_references,bibliography/pns_cooling_references,bibliography/stellarevolution_references,bibliography/methods_references,bibliography/nu_obs_references,bibliography/fluid_dynamics_references}

\end{document}